\documentclass{aa} 

\usepackage{tabularx,caption}
\usepackage{epsfig}
\usepackage[english]{babel}
\usepackage[utf8]{inputenc}
\usepackage{amsmath,amsfonts}
\usepackage{url}

\usepackage{comment}
\usepackage{multirow}

\usepackage[titletoc]{appendix}
\usepackage{natbib}
\usepackage{ulem}

\def \aap{A\&A}
\def \apjl{ApJ}

\def \apjs{ApJS}

\begin{document}

\title{The Gaia-ESO Survey: Catalogue of H$\alpha$ emission stars \thanks{Based on data products from observations made with ESO Telescopes at the La Silla Paranal Observatory under programme ID 188.B-3002. These data products have been processed by the Cambridge Astronomy Survey Unit (CASU) at the Institute of Astronomy, University of Cambridge, and by the FLAMES/UVES reduction team at INAF/Osservatorio Astrofisico di Arcetri. These data have been obtained from the Gaia-ESO Survey Data Archive, prepared and hosted by the Wide Field Astronomy Unit, Institute for Astronomy, University of Edinburgh, which is funded by the UK Science and Technology Facilities Council. This work was partly supported by the European Union FP7 programme through ERC grant number 320360 and by the Leverhulme Trust through grant RPG-2012-541. We acknowledge the support from INAF and Ministero dell' Istruzione, dell' Universit\`a' e della Ricerca (MIUR) in the form of the grant "Premiale VLT 2012". The results presented here benefit from discussions held during the Gaia-ESO workshops and conferences supported by the ESF (European Science Foundation) through the GREAT Research Network Programme. A. Bayo Acknowledges financial support from the Proyecto Fondecyt Iniciaci\'on 11140572. F. Jimenez-Esteban acknowledges financial support from the ARCHES project (7th Framework of the European Union, n 313146).
This research has made use of the SIMBAD database and VizieR catalogue access tool, CDS, Strasbourg, France. The original description of the VizieR service was published in A\&AS 143, 23. Tables \ref{title}, \ref{title2}, and \ref{title3} are only available in electronic form at the CDS via anonymous ftp to cdsarc.u-strasbg.fr (130.79.128.5) or via http://cdsweb.u-strasbg.fr/cgi-bin/qcat?J/A+A/.}
}

\author{G. Traven \inst{1} \and T. Zwitter\inst{1} \and S. Van Eck\inst{2} \and A. Klutsch\inst{3} \and R. Bonito\inst{4,5} \and A.~C. Lanzafame\inst{3,6} \and 
E.~J. Alfaro\inst{7} \and 
A. Bayo\inst{8} \and
A. Bragaglia\inst{9} \and 
M.~T. Costado\inst{7} \and 
F. Damiani\inst{5} \and 
E. Flaccomio\inst{5} \and 
A. Frasca\inst{3} \and 
A. Hourihane\inst{10} \and
F. Jimenez-Esteban\inst{11,12} \and 
C. Lardo\inst{13} \and 
L. Morbidelli\inst{14} \and 
E. Pancino\inst{9,15} \and 
L. Prisinzano\inst{5} \and 
G.~G. Sacco\inst{14} \and
C.~C. Worley\inst{10}
}

\institute{Faculty of Mathematics and Physics, University of Ljubljana, Jadranska 19, 1000 Ljubljana, Slovenia; \\ e-mail: gregor.traven@fmf.uni-lj.si \and 
Institut d'Astronomie et d'Astrophysique, Université Libre de Bruxelles, Boulevard du Triomphe, C.P. 226, B-1050 Bruxelles, Belgium \and 
INAF - Osservatorio Astrofisico di Catania, via S. Sofia 78, 95123, Catania, Italy \and 
Dipartimento di Fisica e Chimica Universita' di Palermo \and 
INAF - Osservatorio Astronomico di Palermo, Piazza del Parlamento 1, 90134, Palermo, Italy \and 
Dipartimento di Fisica e Astronomia, Sezione Astrofisica, Universit\'{a} di Catania, via S. Sofia 78, 95123, Catania, Italy \and 
Instituto de Astrof\'{i}sica de Andaluc\'{i}a-CSIC, Apdo. 3004, 18080 Granada, Spain \and
Departamento de F\'isica y Astronom\'ia, Facultad de Ciencias, Universidad de Valpara\'iso, Av. Gran Breta\~na 1111, 5030 Casilla, Valpara\'iso, Chile \and 
INAF - Osservatorio Astronomico di Bologna, via Ranzani 1, 40127, Bologna, Italy \and 
Institute of Astronomy, University of Cambridge, Madingley Road, Cambridge CB3 0HA, United Kingdom \and
Centro de Astrobiolog\'{\i}a (INTA-CSIC), Departamento de Astrof\'{\i}sica, PO Box 78, E-28691, Villanueva de la Ca\~nada, Madrid, Spain \and 
Suffolk University, Madrid Campus, C/ Valle de la Viña 3, 28003, Madrid, Spain \and 
Astrophysics Research Institute, Liverpool John Moores University, 146 Brownlow Hill, Liverpool L3 5RF, United Kingdom \and 
INAF - Osservatorio Astrofisico di Arcetri, Largo E. Fermi 5, 50125, Florence, Italy \and 
ASI Science Data Center, Via del Politecnico SNC, 00133 Roma, Italy}

\abstract
{We discuss the properties of H$\alpha$ emission stars across the sample of 22035 spectra from the Gaia-ESO Survey internal data release, observed with the GIRAFFE instrument and largely belonging to stars in young open clusters. Automated fits using two independent Gaussian profiles and a third component that accounts for the nebular emission allow us to discern distinct morphological types of H$\alpha$ line profiles with the introduction of a simplified classification scheme. 
All in all we find 3765 stars with intrinsic emission and sort their spectra into eight distinct morphological categories: single--component emission, emission blend, sharp emission peaks, double emission, P-Cygni, inverted P-Cygni, self--absorption, and emission in absorption. We have more than one observation for 1430 stars in our sample, thus allowing a quantitative discussion of the degree of variability of H$\alpha$ emission profiles, which is expected for young, active objects.
We present a catalogue of stars with properties of their H$\alpha$ emission line profiles, morphological classification, analysis of variability with time and the supplementary information from the SIMBAD, VizieR, and ADS databases. The records in SIMBAD indicate the presence of H$\alpha$ emission for roughly 25\% of all stars in our catalogue, while at least 305 of them have already been more thoroughly investigated according to the references in ADS. The most frequently identified morphological categories in our sample of spectra are emission blend (23\%), emission in absorption (22\%), and self--absorption (16\%).
Objects with repeated observations demonstrate that our classification into discrete categories is generally stable through time, but categories P-Cygni and self--absorption seem less stable, which is the consequence of discrete classification rules, as well as of the fundamental change in profile shape.
Such records of emission stars can be valuable for automatic pipelines in large surveys, where it may prove very useful for pinpointing outliers when calculating general stellar properties and elemental abundances. They can be used in studies of star formation processes, interacting binaries, and other fields of stellar physics.}

\keywords{Stars: activity -- Stars: emission-line, Be -- Stars: peculiar -- Galaxy: open clusters and associations: general -- Line: profiles -- Catalogs}

\maketitle

\section{INTRODUCTION}

The Gaia-ESO Survey (GES) is a large spectroscopic survey that targets more than $10^5$ stars in a magnitude range $8 \lesssim V < 19$, systematically covering all major components of the Milky Way, from halo to star forming regions \citep{2012Msngr.147...25G, 2013Msngr.154...47R}. With uniform mode of observation, well-defined samples, based primarily on ESO-VISTA photometry for the field stars, and by a variety of photometric surveys of open clusters, GES will quantify the kinematic-multielement abundance distribution functions of stellar components in the bulge, the thick and thin discs, and the halo, as well as a significant sample of $\sim$ 100 open clusters, covering a wide range of ages, stellar masses, metallicities, and distances from the Galactic centre.

Within the GES, the analysis of different types of stars is divided into different working groups (WGs). WG10 is in charge of the analysis of FGK-type GIRAFFE spectra \citep{2014AA...567A...5R} and WG11 of FGK-type UVES spectra \citep{2014arXiv1409.0568S}. WG12 focuses on sources in the field of young clusters \citep{lanzafame_corrected} and WG13 on OBA-type stars \citep{2015IAUS..307...88B}. WG14 is in charge of non-standard objects, or objects for which standard analyses produce uncertain or no results because of unusual spectroscopic features that are not properly accounted for by automatic pipelines. 

The H$\alpha$ emission, indicative of accretion and outflow activity in young stars, is one of these outstanding features. Indeed, when not properly recognised, it can endanger the parameter or abundance determination analysis, for example because it affects the local continuum placement or the radial velocity determination. It was therefore decided, as part of the WG14 attributions, to systematically detect and characterise H$\alpha$ emission.

This paper presents a catalogue of 4459 H$\alpha$ emission stars and their 8846 spectra across the sample of 22\,035 spectra of 12\,392 stars observed with GIRAFFE instrument from the GES internal data release (first 22 months of observations). Special care is taken to distinguish intrinsic stellar emission from the nebular one. Altogether, we find 7698 spectra with intrinsic emission that belong to 3765 stars. Such stars are often found in stellar nurseries and nebular environments of young open clusters and can therefore serve as a proxy for stellar ages. Another possibility is that they belong to the field and can be so far unrecognised interacting binaries, like symbiotic stars and cataclysmic variables. In this case GES provides a unique census of interacting binaries at all Galactic latitudes, thereby helping to sample the faint luminosity tail of their distribution that is under-represented in existing catalogues.

The H$\alpha$ emission profiles display a variety of shapes depending on the physical mechanism responsible for the emission and on the geometry of the source. The simplest profiles are symmetric with a Gaussian-like appearance and are presumably originated in one distinct source, but other more complex line profiles are frequently observed. We refer the readers to \citet{2013A&A...556A.108B} and \citet{1996A&AS..120..229R} for a discussion of the expected H$\alpha$ profile features and their physical origin. Mechanisms that produce such multicomponent profiles are usually attributed to very young stars, cataclysmic variables, symbiotic stars, stars with massive outflows or inflows, and to many other types of active objects \citep{2012ApJ...751..147H, 2013MNRAS.431.2673K, 2006AA...446.1129B, 2000AA...362..683L}. 

Many attempts have already been made to classify active objects of a certain type with the use of emission lines. An atlas of symbiotic stars was compiled in the context of a H$\alpha$ morphological classification using Gaussian fitting of line profiles \citep{1993A&AS..102..401V}. \citet{2006PASP..118..617R} demonstrate that a double-Gaussian model reproduces the profiles of chromospheric emission lines using a sample of 147 K- and M-type stars. A temporal analysis of Rigel ($\beta$ Ori) spectra showed an example of morphological classification of distinct H$\alpha$ profiles, which the authors classified visually into seven classes \citep{2008cihw.conf..155M}. Another approach at analysis of H$\alpha$ profiles was shown by \citet{2006MNRAS.370..580K} where the authors used radiative  transfer model to describe the various observed profiles of classical T Tauri stars and compared it with the morphological classification scheme proposed by \citet{1996A&AS..120..229R}.

The above studies have mostly dealt with specific types of stars on smaller samples of several dozens of objects. In this paper we do not limit ourselves to any distinct type of stars, as this would require customised treatment when identifying and analysing the very diverse physical conditions purely from observed line profiles. Instead, we try to be impartial towards distinct physical characteristics and treat all available spectra in a uniform manner. In this way, we are able to review and morphologically classify several thousand objects without particular assumptions and model dependent parameters. 

We approach the problem first with the detection of H$\alpha$ emission in spectra and secondly, we introduce a simple morphological scheme for their classification. This classification scheme, together with supplementary information about detected objects, like their classification from the literature where available, might prove useful to those who deal with the aforementioned types of active objects and could serve as a uniform and unbiased treatment of largely diverse emission profiles.

In Sect. 2 we present the data used in this study and the methods of reduction. The results of classification and temporal variability of objects are presented in Sect. 3. Discussion and conclusions follow in Sect. 4.

\section{DATA AND REDUCTION OVERVIEW} \label{sec:data}

\begin{figure*}[!htp]
   \centering
   \includegraphics[width=\textwidth]{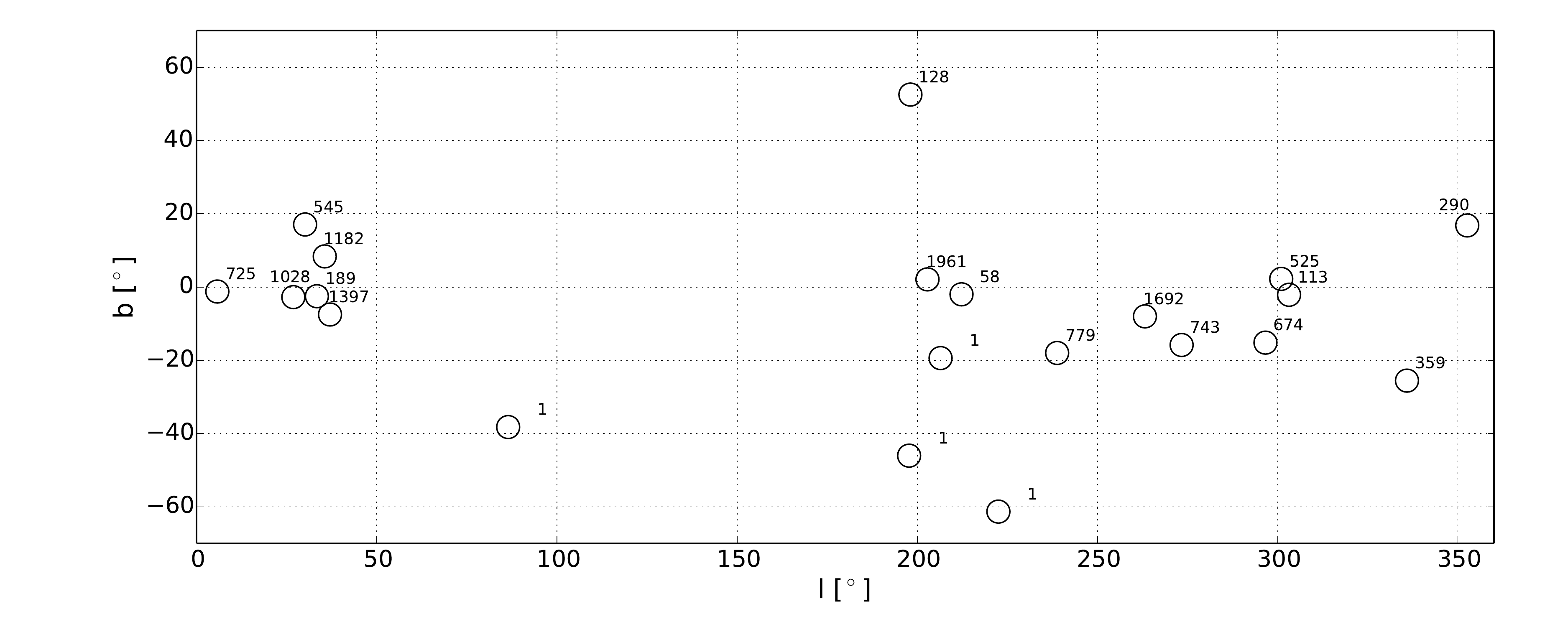}                        
   \caption{Whole sky representation of the 20 fields marked by circles. At the top right of each circle, we indicate the number of objects within a given field, where one-object fields represent benchmark stars.}
\label{targets}
\end{figure*}

The GES employs the VLT-UT2 FLAMES high-resolution multiobject spectrograph  \citep{2002Msngr.110....1P}. FLAMES has two instruments, the higher-resolution  UVES, fed by 8 fibres, and GIRAFFE,  with about 130 fibres. In the present paper we concentrate on the much more numerous GIRAFFE spectra to thoroughly test our classification scheme;  a similar analysis of the UVES spectra will follow later.

We use only spectra obtained in the wavelength range with H$\alpha$ line - HR15N setting (6470--6790 \AA\,, R=17\,000), for which the majority of targets selected by GES are open cluster stars. Along with ongoing GES observations, we also use all relevant ESO archive data, which are included to be analysed as part of GES. Individual spectra can therefore be dated up to nine years before the start of the survey. For objects with more than one spectrum from either new or archival data, the timespan of repeated observations ranges from two days up to one year, which enables us to consider their temporal variability.

In this study, we use only spectra before sky subtraction and normalisation. Doing so, we avoid any wrong identification or inaccurate subtraction of nebular emission lines, which may show a significant spatial variability in the environment of young clusters. Out of altogether 28\,957 spectra from the GES internal data release, we have 22\,035 object spectra (i.e. 76\% of the whole data) while the remaining 6922 spectra are measurements of the background sky signal. The distribution of targets on the sky is shown in Fig. \ref{targets}.  

Next we describe the methods used to characterise spectral H$\alpha$ emission. Here we give just a brief overview, and more details are given in Appendix \ref{sec:detect}. 

The entire analysis and the final results are based on spectra reduced to the heliocentric frame of reference. Normalisation was done by dividing the spectra with the median value of the flux in a window of width 20 {\AA} around the H$\alpha$ line. In the first step of the line profile analysis, we determinated the separation of absorption-dominated and emission-dominated profiles. This was done by analysing the steepness and strength of the left and right part of the line profile. Out of 22\,035 spectra, we identified 9265 emission profiles, taking all those with one or more H$\alpha$ emission components into account. Some of these spectra represent repeated observations of the same star so the number of unique emission type objects is 4566 out of 12\,392.

\begin{figure*}[!htp]
   \centering
   \includegraphics[trim = 18mm 0 18mm 0, clip, width=0.32\textwidth]{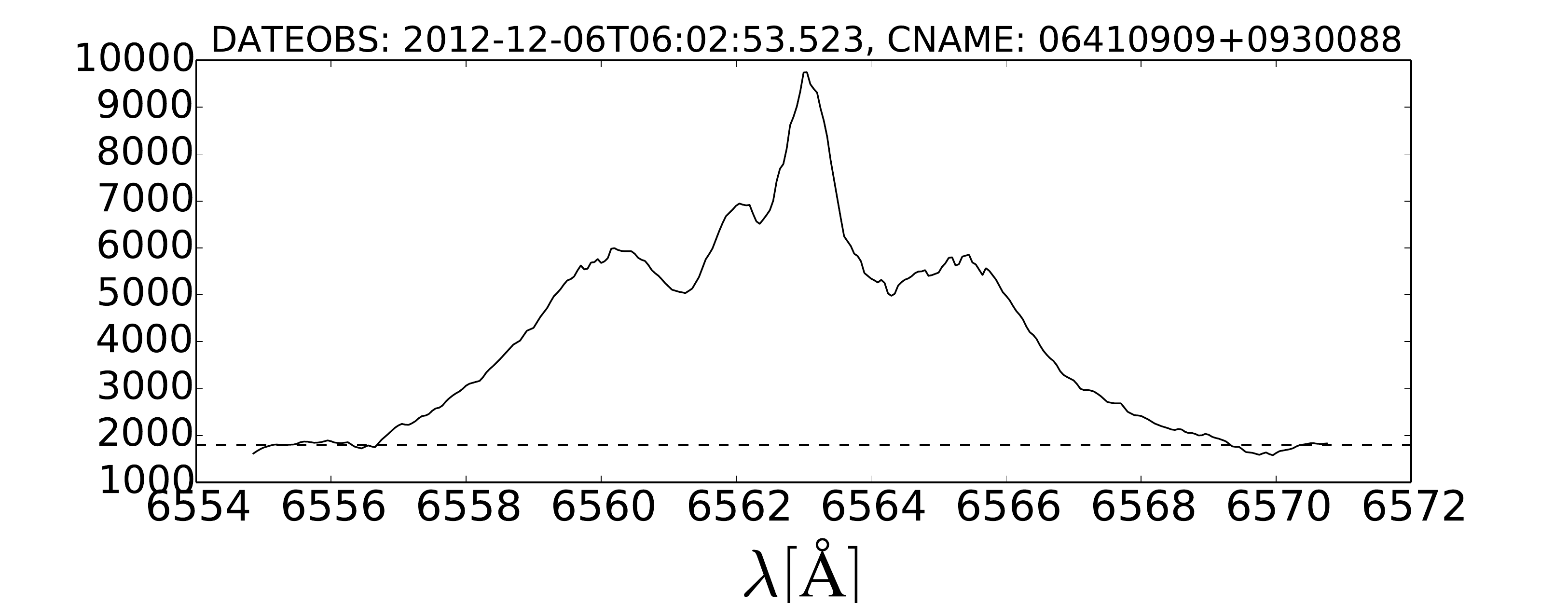} 
   \includegraphics[trim = 18mm 0 18mm 0, clip, width=0.32\textwidth]{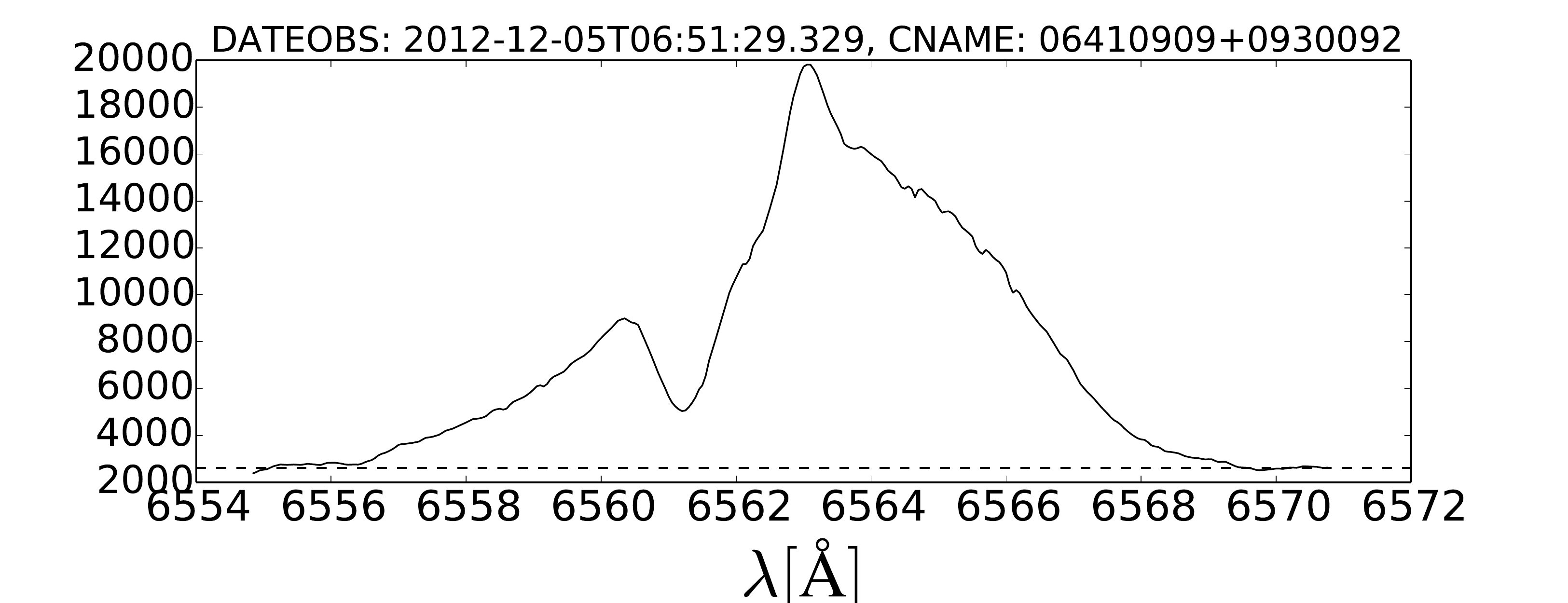}
   \includegraphics[trim = 18mm 0 18mm 0, clip, width=0.32\textwidth]{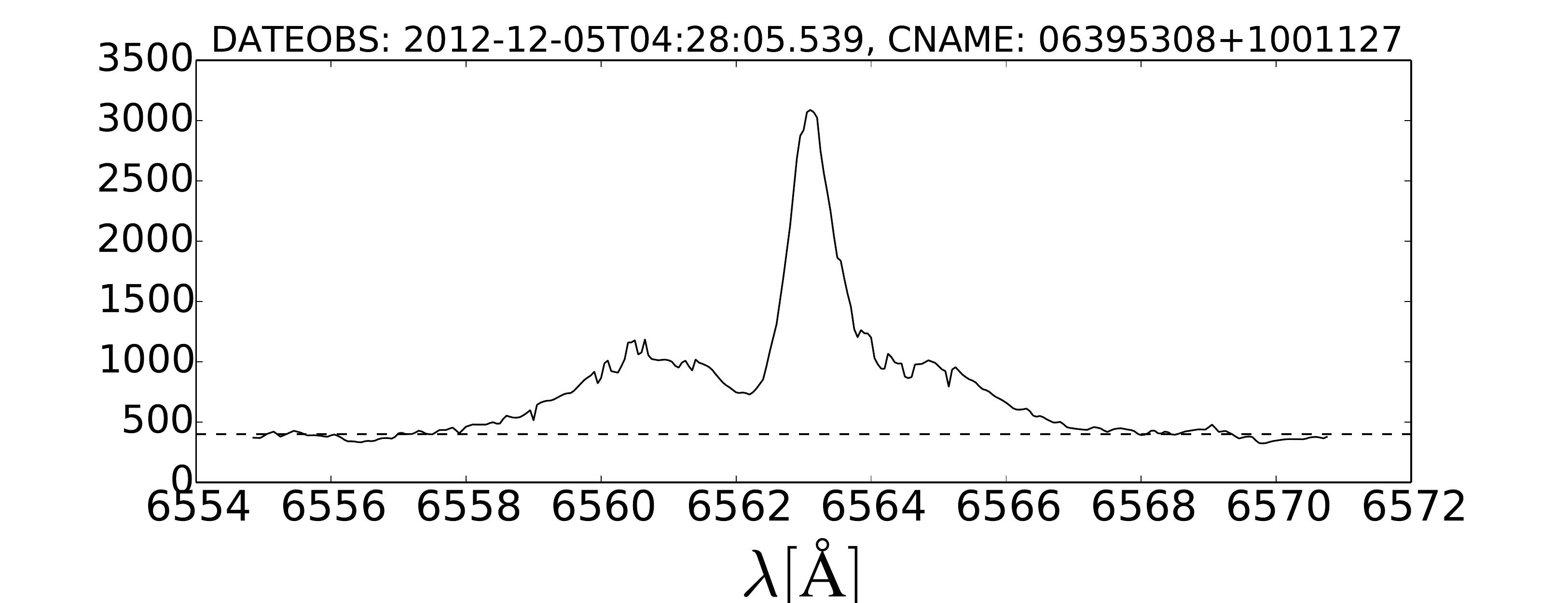}
   \includegraphics[trim = 18mm 0 18mm 0, clip, width=0.32\textwidth]{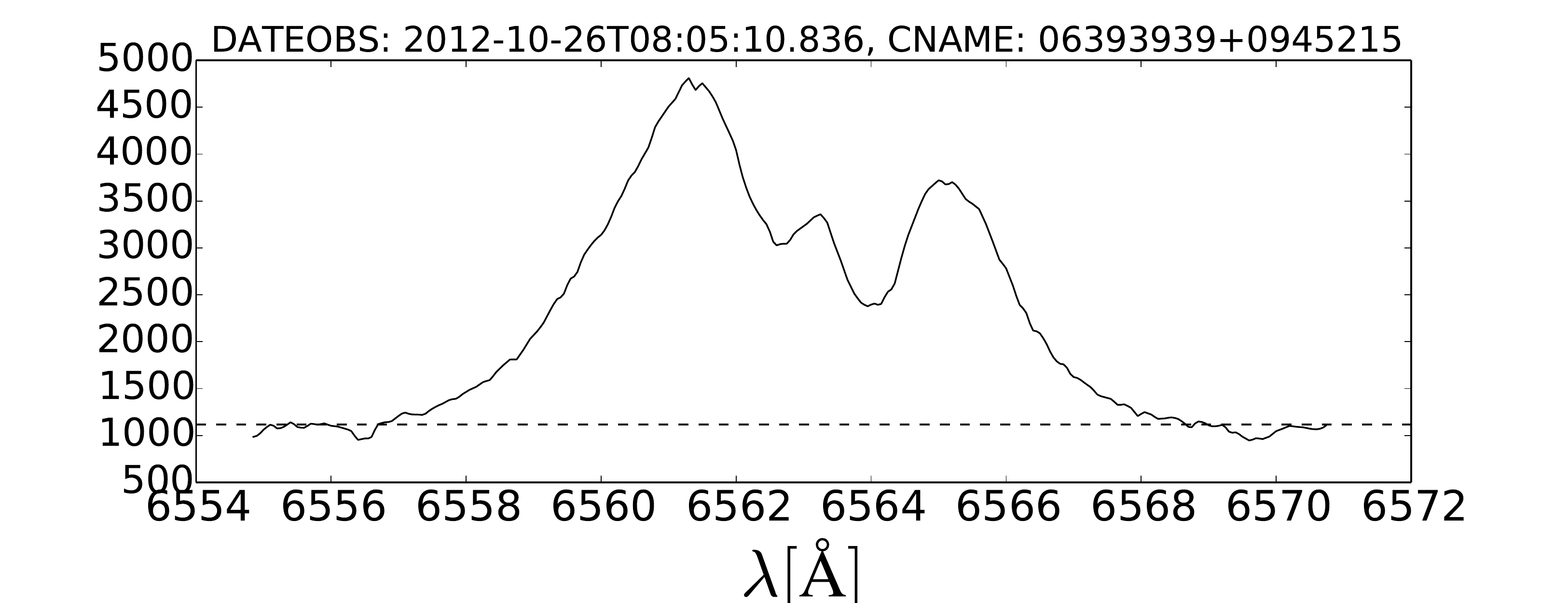}
   \includegraphics[trim = 18mm 0 18mm 0, clip, width=0.32\textwidth]{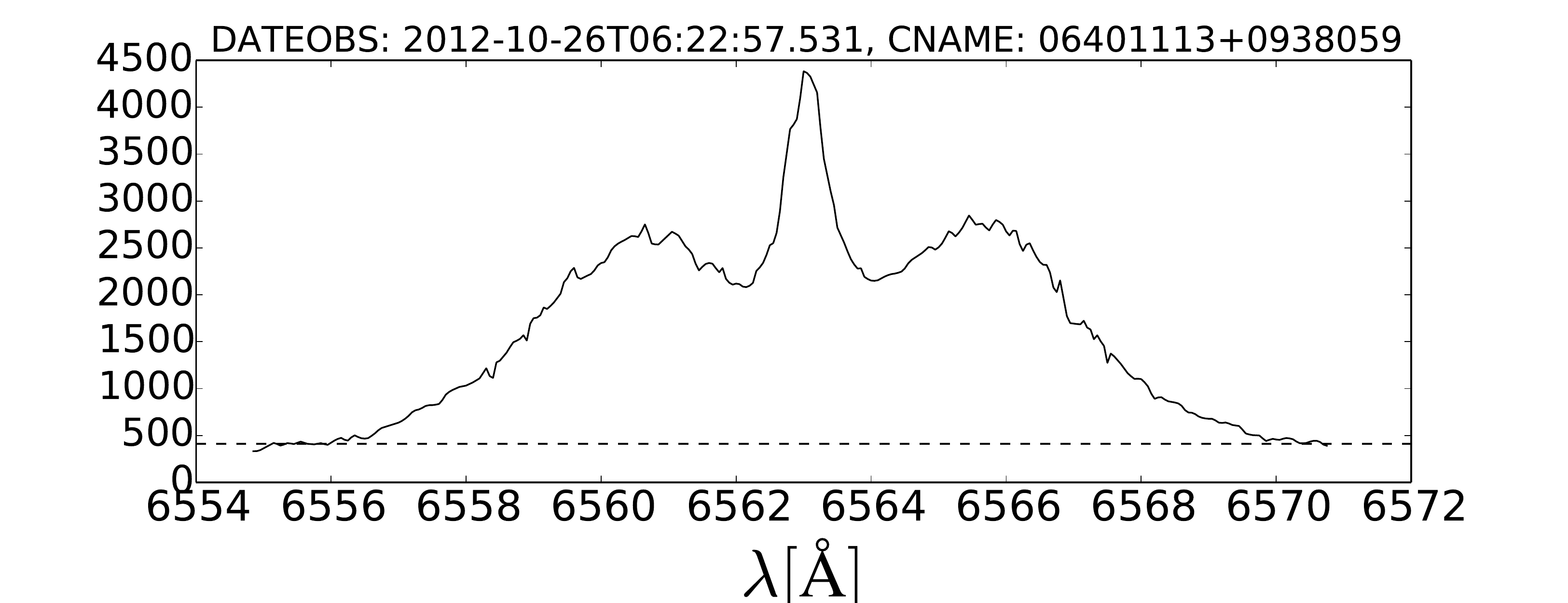}
   \includegraphics[trim = 18mm 0 18mm 0, clip, width=0.32\textwidth]{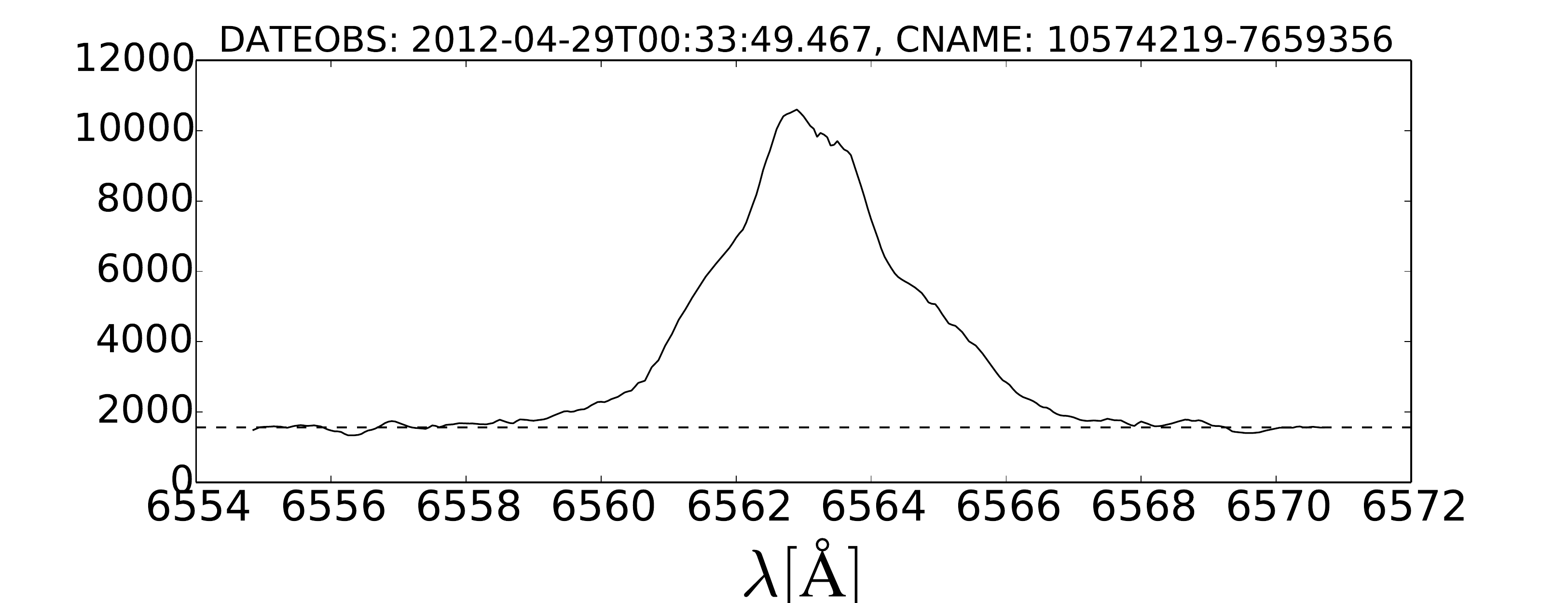}                      
   
   \caption{Examples of GES spectra with H$\alpha$ in emission before normalisation and sky subtraction, plotted in heliocentric wavelengths. Dashed lines mark the adopted continuum level.}
   \label{spectra}
\end{figure*}

Both simple and more complicated multicomponent profiles are present in our dataset (Fig. \ref{spectra}). The goal of further examination was to determine the number of separate emission or absorption components in each profile and whether a nebular component is present. Clearly a solution with more components should always yield a better fit. But blending of individual components often makes the solution degenerate, while large intensity ratios make the claim of multiple component detection insignificant. To avoid degeneracy and to increase uniformity of treatment of a wide variety of line profiles, we decided to fit each profile by the combination of two independent Gaussian profiles either in emission or absorption, hereafter denoted as $g_1$ and $g_2$, and an optional third one, denoted $g_n$, to account for a nebular emission. We assume H$\alpha$ as the source of all components, neglecting possible contribution from [He~II] absorption line (6560.15 {\AA}) in hot stars. 

Each Gaussian profile has three parameters for which we use a standard notation ($\lambda_i$ - centre of the peak, $\sigma_i$ - standard deviation, $A_i$ - height of the peak, $i = {1,2,n}$). Resorting to one Gaussian profile is not enough to describe diverse multicomponent profiles, while 3 or more are plagued by degeneracy. The two component fit on the other hand describes well the emissions with self--absorption, P-Cygni or inverted P-Cygni profiles, or narrow emission peaks within a dominating absorption. The optional third (nebular) component is included in the fit if there is strong indication of the nebular emission. This indication comes from the presence of [N~II] emission lines in the object spectrum or from the presence of H$\alpha$ emission line in the sky spectra from the region surrounding the object. In both cases, the nebular component is assumed to be in emission, where its wavelength and width are fixed by [N~II] lines and H$\alpha$ line in the sky spectra, as explained in Appendix \ref{gauss_fitting}.

Our analysis is based exclusively on automated fits. This should improve the uniformity of the results and avoid arbitrary decisions required in manual fitting. Automated multicomponent fits can be computationally intensive with results reflecting local minima of the solution, thus missing the global one. To avoid these problems we adopt a sampling scheme for all the Gaussian parameters except intensities, which are fitted with a least-squares Levenberg–Marquardt algorithm. We fix the $\lambda_{n}$ and $\sigma_{n}$ values for $g_n$ based on the evaluation of H$\alpha$ and [N~II] emission, while we leave the peak height ($A_n$) to vary in the fit. The $g_n$ component is included in the fit only if it satisfies criteria, as detailed in Appendix \ref{gauss_fitting}. We sample the $\sigma$ parameter of $g_1$ and $g_2$ (hereafter $\sigma_{1,2}$) logarithmically with 15 values ranging from 0.15 to 2 {\AA}. The $\lambda$ parameter of $g_1$ and $g_2$ (hereafter $\lambda_{1,2}$) are, in the first iteration of the fit, distributed linearly with 21 values on a flexible interval around the detected centre of H$\alpha$ emission. Justification of the choice of these intervals is given in Appendix \ref{gauss_fitting}. The fit is then iterated so that the final precision for the $\lambda_{1,2}$ parameters reaches 0.1 {\AA}. At this point we assume that the $\lambda_{1,2}$ values from the first iteration are close to the global minimum, so we use a better sampling to investigate the parameter space only around these values. The peak heights $A_{1,2}$ of $g_1$ and $g_2$ are allowed to vary continuously, as for $g_n$. 

The raw results of this fitting procedure are 9 or 6 Gaussian parameters ($\lambda_{1,2,n}, \sigma_{1,2,n}, A_{1,2,n}$) for 3 or 2 fitted Gaussian components respectively along with the value of the goodness of fit defined as
\begin{equation} \label{chi2}
\chi^2_r = \left(\sum_i^N [O-C]_i^2 \right) / (N - N_{var})
\end{equation}
where $N$ is the number of data points and $N_{var}$ is the number of input parameters to the least-squares method. We prefer not to use the expected $\sigma$ (dispersion) of the noise based on the S/N (signal-to-noise ratio) of the spectra since it might be influenced by systematic errors (e.g. uncertain placement of continuum, imperfect flat fielding) and statistical (Poisson) noise. Because it is physically plausible that the central wavelength of emission and absorption profiles are the same, or that an emission profile has a narrow core and broad wings, which cannot be fitted with a single Gaussian, we also perform this fitting procedure by setting $\lambda_1 = \lambda_2$. By doing this, we anticipate to get better solutions of the fit in the case where two components in line profiles are well aligned and the original procedure misses the right initial values of $\lambda$ to have the wavelengths match exactly. This version of fitting produces better fits in terms of $\chi^2_r$ for 402 spectra. As these solutions are also meaningful in the physical sense we adopt these new values.\\

\section{RESULTS}

The results from the fitting procedure for 8846 out of 9265 spectra are contained in the catalogue described in Appendix \ref{sec:contents}. We disregard 396 spectra due to the strong negative values of flux around H$\alpha$ in the spectra before normalisation (see Appendix \ref{sec:detect}). We also disregard 23 (i.e.\ less than 0.3\% of all) spectra with absorption profiles, which were misidentified as emissions in the first step of the data reduction (e.g. due to noise or cosmic rays). In the presentation of results, the velocity scale is often used for the Gaussian parameter $\lambda$ (thus denoted by $v$), with zero at the H$\alpha$ centre.

\subsection{Stars with repeated observations}

The fitted sample of 8846 spectra represents data for 4459 unique objects. We therefore have 3029 objects in the catalogue represented with one spectrum and 1430 objects with repeated observations. The latter can be used to illustrate the temporal variability of H$\alpha$ line profiles. They are also used for determination of the free parameters in our morphological classification scheme. The number of spectra for each of the 1430 objects varies from 2 to 21 (Table \ref{title5}). The GES introduced an observing strategy \citep{bragaglia}, which enables such existence of multiple spectra per star. Many of these spectra are also from the ESO archive data.

\begin{table}
\captionof{table}{Frequency of multiple observations. The majority of objects (3029) have been observed once. Among the 1430 sources observed several times, 120 have over ten observations.} 
\normalsize
\centering
\begin{tabular}{lllllllllllllllllllllll}
\bf N spec. & 1&2&3&4&5&6&7&8\\
\hline
\bf N obj. & 3029 & 784 & 295 & 149 & 55 & 17 & 9 & 1 
\vspace{0.3cm}
\end{tabular}

\begin{tabular}{lllllllllllllllllllllll}
\bf N spec. & 9&11&12&15&16&18&19&20&21\\
\hline
\bf N obj. & 1 & 2 & 2 & 1& 4& 2& 14 &90 &4

\end{tabular}
\bigskip
\label{title5} 
\end{table}

\subsection{Morphological classification}

Multiple Gaussian profile fit described in the previous section is a morphological one, so it is not necessarily related to the underlying physics of observed emission object. We deliberately choose a simple scheme using only up to two Gaussian profiles to describe the H$\alpha$\ line shape intrinsic to the object. Hence we can use a limited number of parameters to classify all the 4459 analysed objects with their 8846 spectra by meaningful morphological categories and possibly also by physical ones.

Classification is enabled by identification of the fitted components described in Appendix \ref{sec:identification}, which produces nine distinct flags for all emission-type objects. Employing this identification, we now focus on the more interesting profiles with two intrinsic components flagged EE(N), EA(N), and AE(N). Letters E, A, and N denote intrinsic emission, intrinsic absorption, and nebular emission, respectively. Flag EA represents a profile where emission is relatively stronger than absorption, and the opposite is true for AE. For more information on flags see Appendix \ref{sec:identification}. 

We establish seven classification categories to infer some correspondence to underlying physics for these spectra (objects). The proposed seven categories with the additional two categories for nebular and single--component intrinsic emission profiles are listed in Table \ref{phys_cat} together with conditions for their assignment. This classification scheme uses three parameters ($K_{EE}$, $K_{Cyg}, K_{sp}$), which are described later in this section, and their values are determined with the help of repeated observations. The latter is done using only 1191 objects for which we have at least two spectra flagged as EE(N), EA(N), or AE(N). 

In Table \ref{classbyclass} we sort 4443 spectra of 1191 objects in rows and columns according to the scheme in Table \ref{phys_cat}. Each object contributes the number of its spectra (categories) into the row that corresponds to the most frequent (prevalent) category for all spectra of that object. If several (N) categories are assigned the same (highest) number of spectra for a given object, then that object contributes to the corresponding N rows, and the number of spectra with a certain category is divided by N before being added to the columns. Finally, values in each row are converted to fractions of the row's sum. Diagonal values indicate the stability of classification categories with time. We note that the table is not diagonally symmetrical, which would be the case if all objects had two spectra only. Here however, the category that is in minority for a certain object contributes to the off-diagonal elements while the prevalent category does not, except where prevalence is shared by more than one category. 

The parameters $K_{EE} = 0.9$, $K_{Cyg} = 0.9$ and $K_{sp} = 50$ km s$^{-1}$ are selected so that they maximise the overall stability of the classification categories. It is physically and morphologically sensible that we relate the degree of separation of two components to a function of the components' widths ($\sqrt{\sigma_1^2 + \sigma_2^2}$), whereby we add an arbitrary constant, which should be close to 1. This constant is then determined by the stability of categories with time, which is indicated by the diagonal elements in Table \ref{classbyclass}. Our conditions from Table \ref{phys_cat} and the parameters $K_{EE}, K_{Cyg}$, and $K_{sp}$ are more supportive of the classification scheme if the diagonal elements are high for all categories. Conversely, if we get many significant off-diagonal elements, this might point to the actual morphological change of profiles for certain objects. We especially expect the latter from 108 objects with at least 10 spectra (Table \ref{title5}), which might be targeted through their variable nature.

\begin{table*}
\captionof{table}{Classification categories and corresponding conditions.}
\footnotesize
\centering
\begin{tabular}{ r c c c c c }
\bf Category & \bf Flags & \bf Condition 1 & \bf Condition 2 & \bf N spectra & \bf Fraction\\
\hline
Nebular emission & N &  &  & 1148 & 13 \%\\
Single--component emission & E(N) &  &  & 990 & 11.2 \%\\
Emission blend ({\bf E$_{bl}$}) & EE(N) & $|\lambda_1 - \lambda_2| \leqslant K_{EE} (\sigma_1^2 + \sigma_2^2)^{1/2}$  &  & 1729 & 19.6 \%\\
Sharp emission peaks ({\bf E$_{sp}$}) & EE(N) & $|\lambda_1 - \lambda_2| > K_{EE} (\sigma_1^2 + \sigma_2^2)^{1/2}$ & $|v_1 - v_2| < K_{sp}$ & 765 & 8.6 \%\\
Double emission ({\bf E$_{dp}$}) & EE(N) & $|\lambda_1 - \lambda_2| > K_{EE} (\sigma_1^2 + \sigma_2^2)^{1/2}$ & $|v_1 - v_2| \geqslant K_{sp}$ & 545 & 6.2 \%\\
P-Cygni ({\bf P$_{Cyg}$}) & EA(N), AE(N) & $\lambda_{Em} - \lambda_{Abs} > K_{Cyg} (\sigma_{Em}^2 + \sigma_{Abs}^2)^{1/2}$ &  & 154 & 1.8 \%\\
Inverted P-Cygni ({\bf IP$_{Cyg}$}) & EA(N), AE(N) & $\lambda_{Abs} - \lambda_{Em} \geqslant K_{Cyg} (\sigma_{Em}^2 + \sigma_{Abs}^2)^{1/2}$ &  & 610 & 6.9 \%\\
Self--absorption ({\bf S$_{abs}$}) & EA(N), AE(N) & $|\lambda_{Abs} - \lambda_{Em}| \leqslant K_{Cyg} (\sigma_{Em}^2 + \sigma_{Abs}^2)^{1/2}$ & $\sigma_{Abs} < \sigma_{Em}$ & 1253 & 14.2 \%\\
Emission in absorption ({\bf E$_{abs}$}) & EA(N), AE(N) & $|\lambda_{Abs} - \lambda_{Em}| \leqslant K_{Cyg} (\sigma_{Em}^2 + \sigma_{Abs}^2)^{1/2}$ & $\sigma_{Abs} \geqslant \sigma_{Em}$ & 1652 & 18.7 \%\\
\hline
&&&& 8846 & 100 \%
\end{tabular}
\label{phys_cat}  
\end{table*}

\begin{table}
\captionof{table}{Fractions of categorisations of the same object.}
\footnotesize
\centering
\begin{tabular}{ c c c c c c c c }
prevalent  & \multicolumn{7}{c}{ Categories of spectra of the same object }\\
category & {\bf E$_{bl}$} & {\bf E$_{sp}$} & {\bf E$_{dp}$} & {\bf P$_{Cyg}$} & {\bf IP$_{Cyg}$} & {\bf S$_{abs}$} & {\bf E$_{abs}$}\\
\hline
{\bf E$_{bl}$} & \bf 80.2 & 0.8 & 3.0 & 0.4 & 0.5 & 13.5 & 1.5\\
{\bf E$_{sp}$} & 1.4 & \bf 80.0 & 0.1 & 0.5 & 4.0 & 3.6 & 10.4\\
{\bf E$_{dp}$} & 4.0 & 0.1 & \bf 73.6 & 1.7 & 1.3 & 18.6 & 0.6\\
{\bf P$_{Cyg}$} & 1.9 & 9.7 & 1.0 & \bf 64.0 & 3.5 & 3.3 & 16.7\\
{\bf IP$_{Cyg}$} & 1.4 & 5.2 & 0.1 & 0.2 & \bf 77.2 & 5.3 & 10.5\\
{\bf S$_{abs}$} & 12.1 & 2.5 & 13.2 & 1.5 & 1.8 & \bf 67.1 & 1.9\\
{\bf E$_{abs}$} & 0.3 & 4.3 & 0.1 & 1.7 & 5.3 & 1.6 & \bf 86.7\\

\end{tabular}
\label{classbyclass}  
\end{table}

Following the scheme in Table \ref{phys_cat} we assign a classification category to all 8846 spectra. The last two columns in Table \ref{phys_cat} give an overview of the number of objects that belong to each category. The fractions in the last column are merely indicating the properties of our sample and are not intended to demonstrate any general population characteristics. 

We observe two subtypes in the E flagged (single--component emission) spectra. The first one consists of very narrow profiles where the mean value of the $\sigma$ parameter of fitted Gaussian profiles is at $\sim 0.29$ {\AA}. This could indicate the threshold for the width of the profile above which we are able to discern distinct components, and can serve as a confirmation of the high quality/resolution of GES spectra. The second subtype corresponds to Gaussian profiles with much larger widths (up to 2 {\AA}), which are also relatively strong, so that additional components are discarded by the ruleset from Fig. \ref{class_scheme}.

\subsubsection{Description of morphological categories}

\begin{figure}[!htp]
   \centering
   \includegraphics[width=\linewidth]{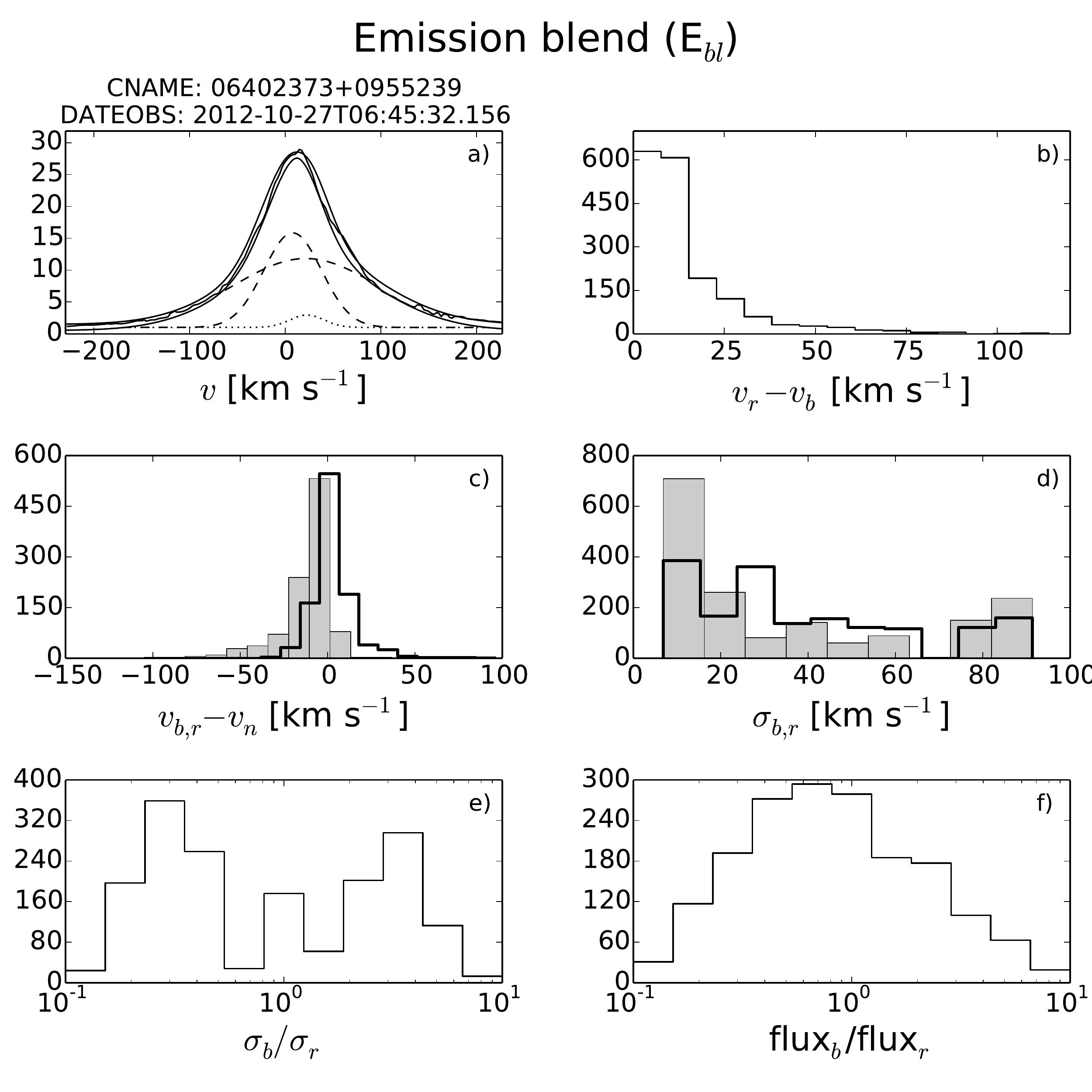}                        
   \caption{Category {\bf E$_{bl}$}. Panel (a) represents the typical continuum-normalised H$\alpha$ profile for objects in this category (a spectrum without detection of prominent nebular emission lines was chosen for clarity). We also plot each Gaussian fit ($g_1, g_2$ - dashed, $g_n$ - dotted) along with their sum (thick black-white line). The horizontal axis is centered on H$\alpha$ at 6562.8 {\AA}. The panels (b) to (f) show the distribution of fitted parameters for all the sources of this category. Parameters for the nebular component are only shown in panel (c) as a reference point for zero wavelength and the number of spectra is therefore different for that histogram, since not all spectra have the additional N flag. Indices $b, r, A, E$ denote blue (shaded histogram), red (thick-lined histogram), absorption (shaded histogram), and emission (thick-lined histogram) component, respectively.}
   \label{Em_blend}
\end{figure}

\begin{figure}[!htp]
   \centering
   \includegraphics[width=\linewidth]{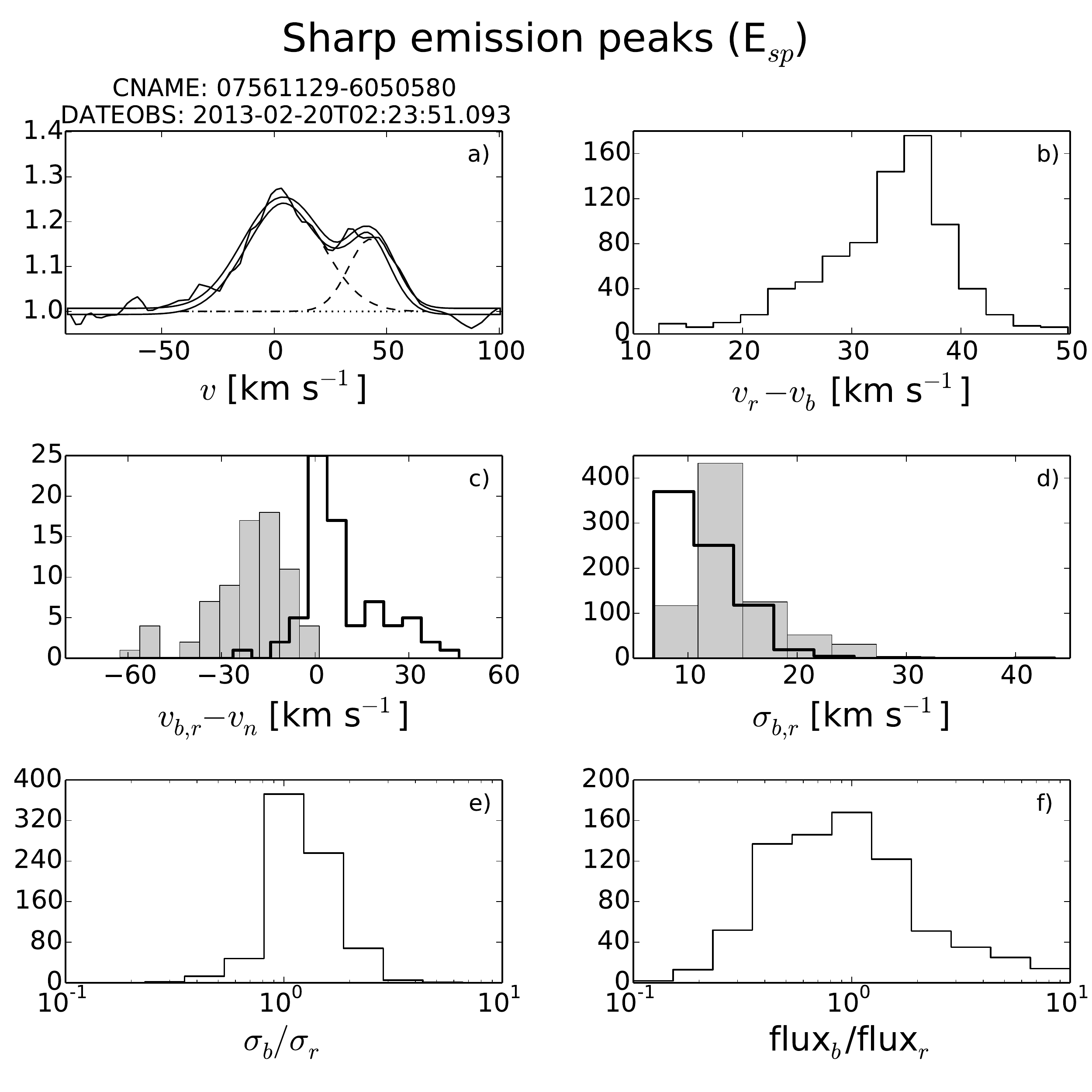}                        
   \caption{As Fig. \ref{Em_blend} but for category {\bf E$_{sp}$}.}
   \label{sharp-peak}
\end{figure}

\begin{figure}[!htp]
   \centering
   \includegraphics[width=\linewidth]{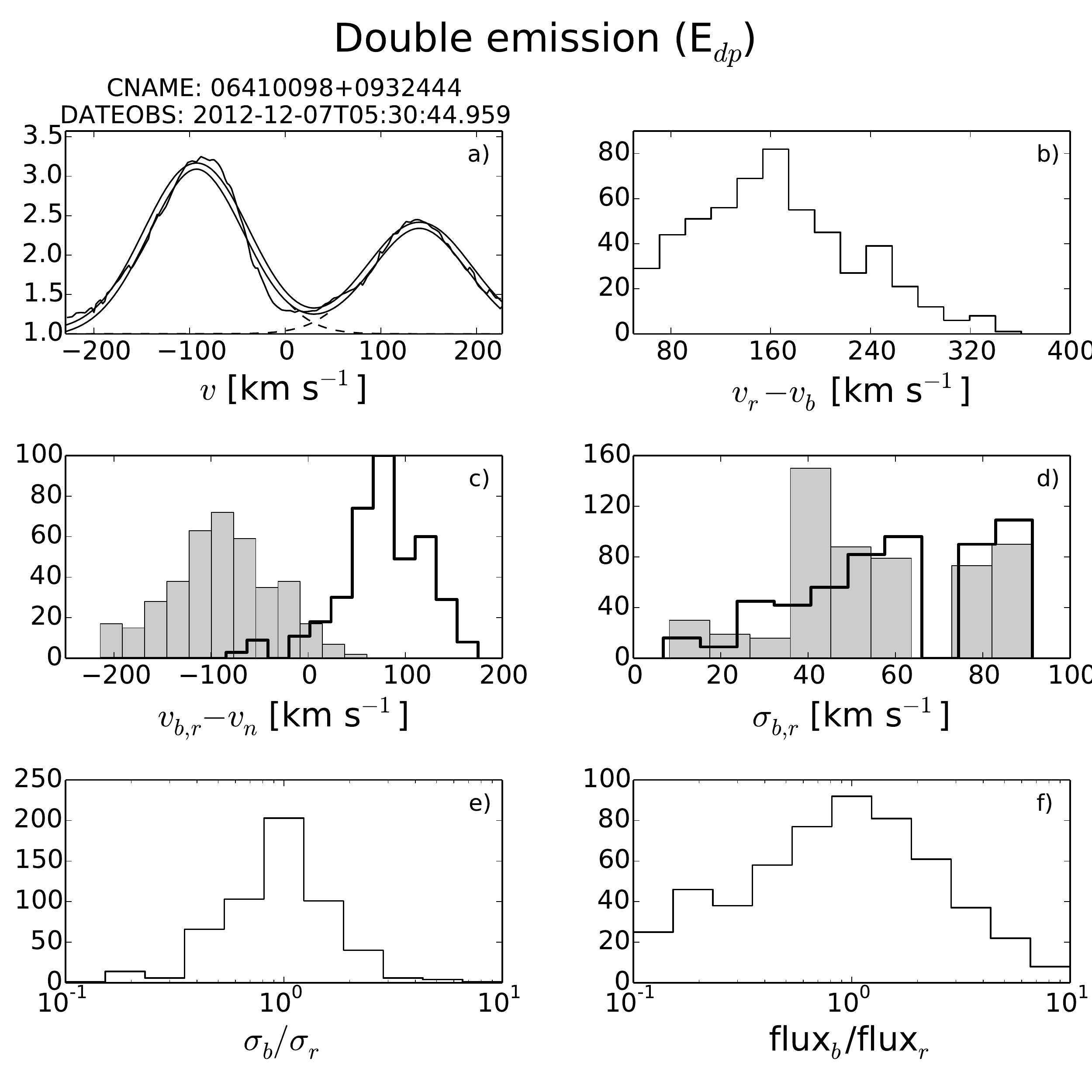}                        
   \caption{As Fig. \ref{Em_blend} but for category {\bf E$_{dp}$}.}
   \label{double_Em}
\end{figure}

\begin{figure}[!htp]
   \centering
   \includegraphics[width=\linewidth]{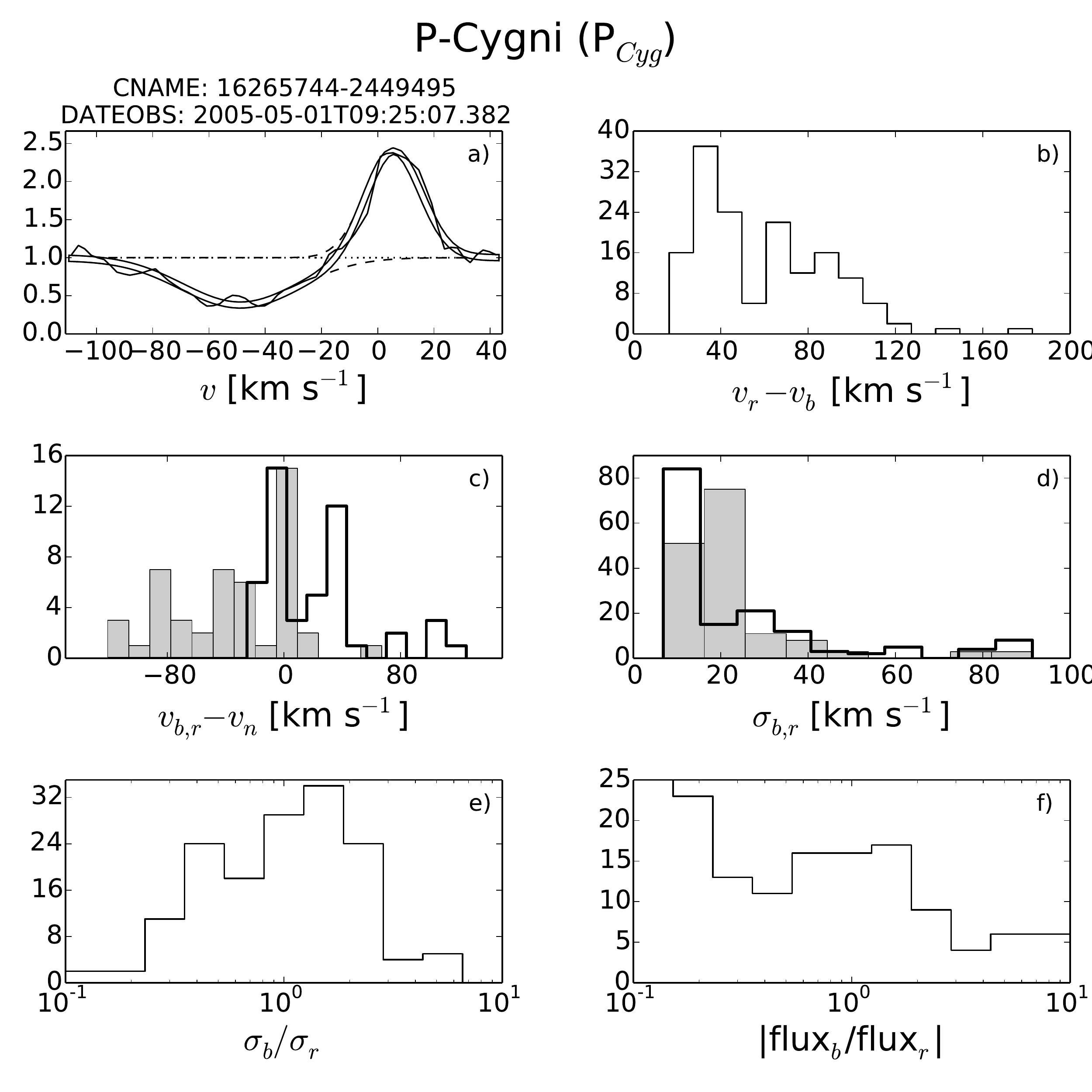}                        
   \caption{As Fig. \ref{Em_blend} but for category {\bf P$_{Cyg}$}.}
   \label{P-Cygni}
\end{figure}

\begin{figure}[!htp]
   \centering
   \includegraphics[width=\linewidth]{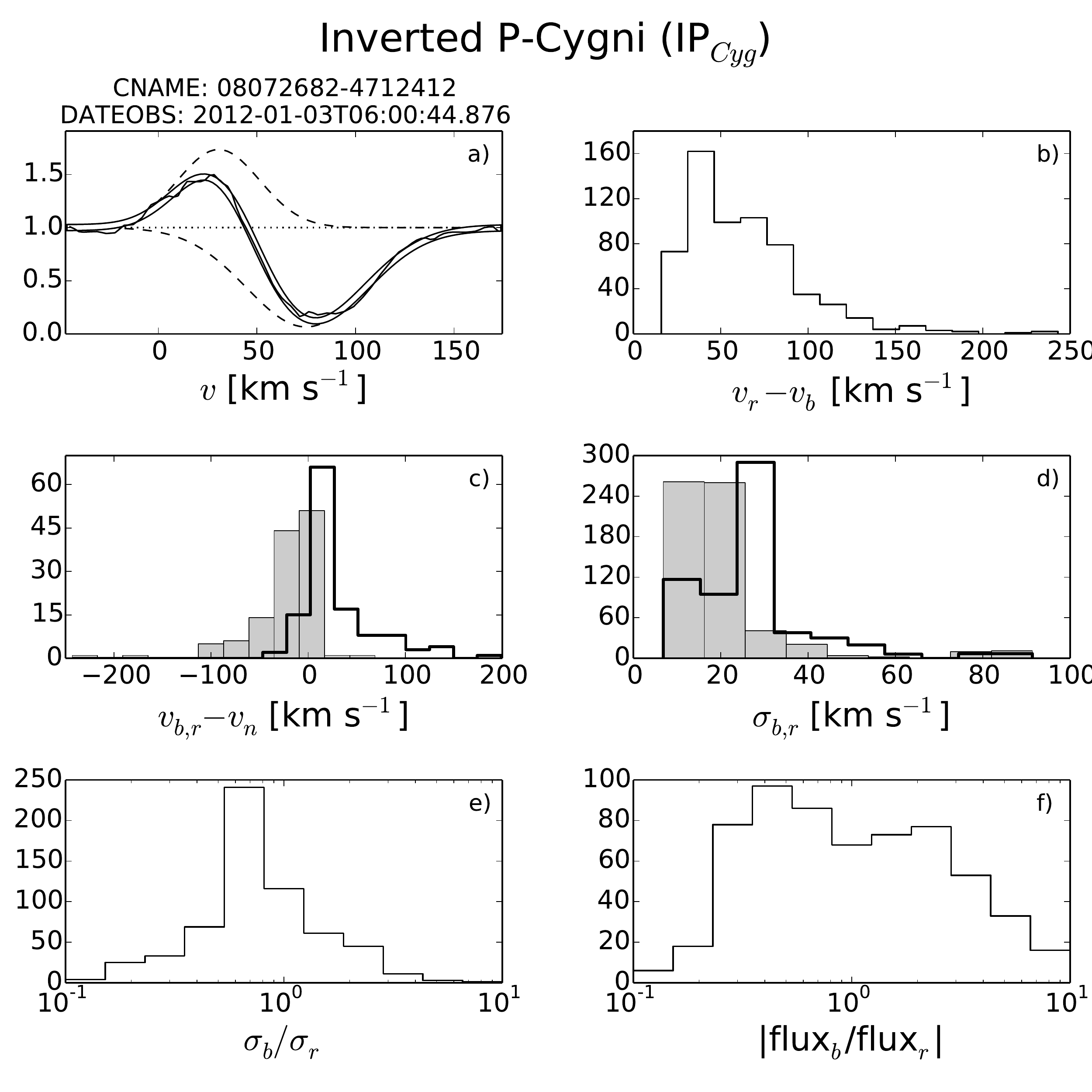}                        
   \caption{As Fig. \ref{Em_blend} but for category {\bf IP$_{Cyg}$}.}
   \label{InvP-Cygni}
\end{figure}

\begin{figure}[!htp]
   \centering
   \includegraphics[width=\linewidth]{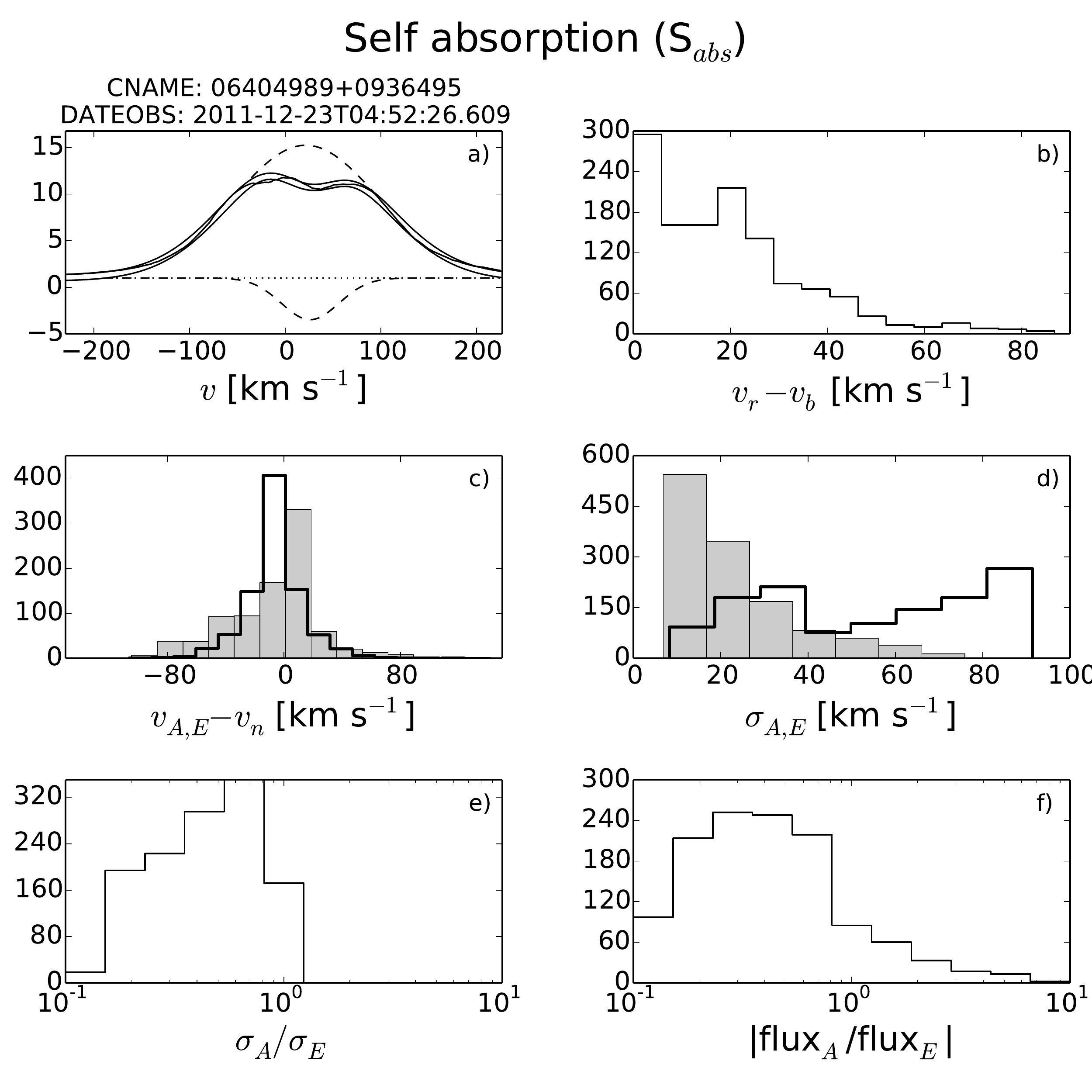}                        
   \caption{As Fig. \ref{Em_blend} but for category {\bf S$_{abs}$}.}
   \label{Self-Abs}
\end{figure}

\begin{figure}[!htp]
   \centering
   \includegraphics[width=\linewidth]{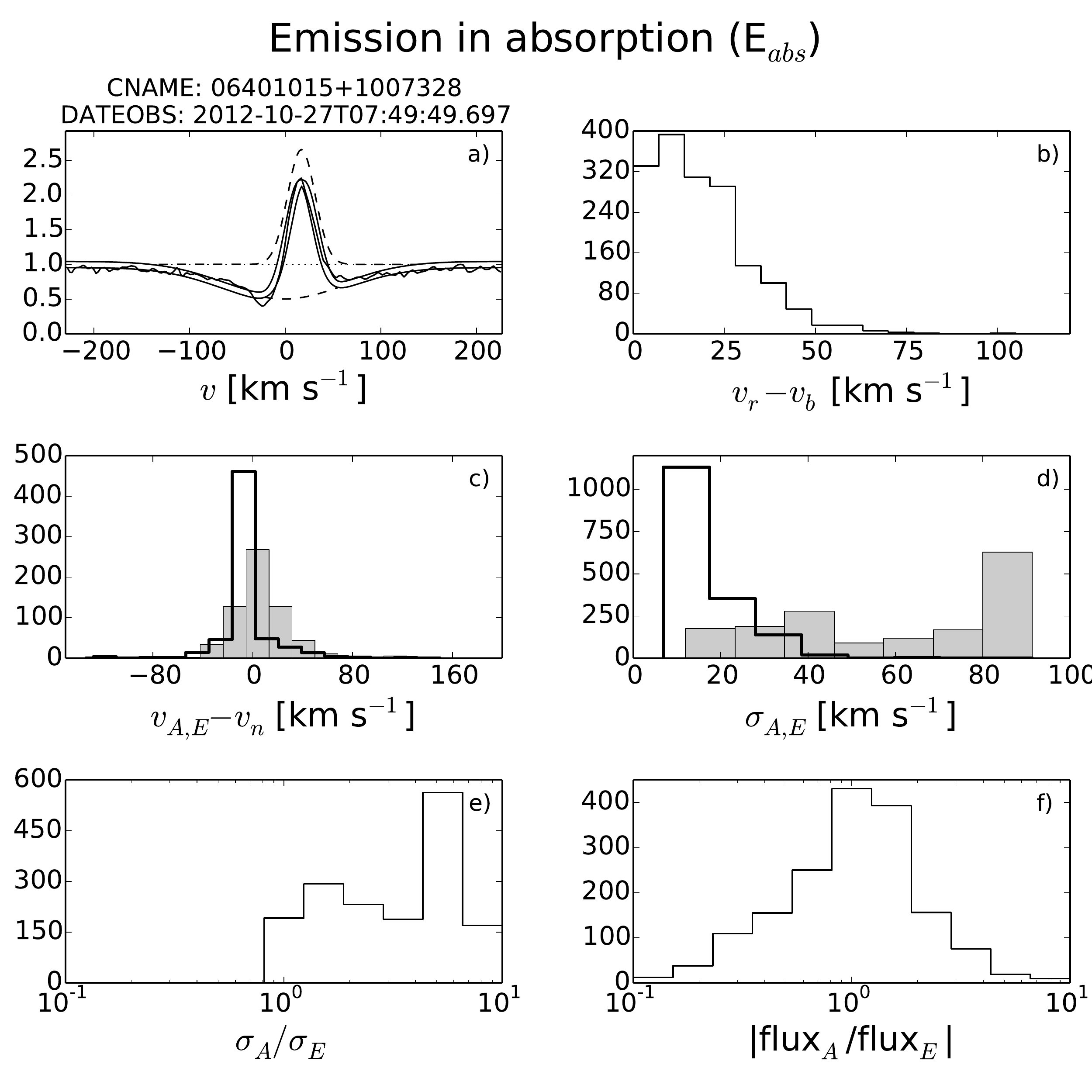}                        
   \caption{As Fig. \ref{Em_blend} but for category {\bf E$_{abs}$}.}
   \label{Em_in_Abs}
\end{figure}

We describe the ensemble properties of spectra for each of the categories using Figs. \ref{Em_blend} - \ref{Em_in_Abs}. All figures are composed of the same number and type of panels, although the reader must be cautious when comparing different categories because of the different ranges on the horizontal axes. The different types of histograms illustrate the qualitative (shape) and quantitative differences between the classification categories based on the parameters of $g_1$ and $g_2$ (denoted also with indexes $b, r, A, E$ for bluer, redder, absorption, and emission component, respectively).  

Category {\bf E$_{bl}$} represents profiles with blended $g_1$ and $g_2$. We observe the slowly falling wing of the profile on either side. Widths of both components can be very different. The redder component is normally centred on nebular velocity, which might also be a consequence of the nebular emission and the presence or lack of the correspondingly fitted $g_n$ component. The bluer component seems to be weaker and narrower on average and its shift from the systemic (nebular) velocity might be due to intrinsic properties (e.g. an approaching wind).

Category {\bf E$_{sp}$} describes profiles with peaks at separations between 20 and 40 km s$^{-1}$. The redder and sharper component is normally centred on nebular velocity, which could be due to fitting of the nebular component with both $g_n$ and $g_r$, in which case only the bluer component is intrinsic. 

Profiles with larger peak separations (between 100 and 250 km s$^{-1}$) are characterised by {\bf E$_{dp}$} category. Widths are between 25 and 100 km s$^{-1}$ so the peaks are well separated. Peaks can be very different, but distributions of width ($\sigma_{b, r}$) and flux ratios are symmetric.

Category {\bf P$_{Cyg}$} accounts for profiles with absorption and emission components that are similar to P-Cygni profiles. The redder emission peak is roughly at systemic (nebular) velocity. Absorption component is usually weaker, which is a common feature for this type of profiles.

Spectra with profile shape features that are generally inverse to those of P-Cygni are represented by {\bf IP$_{Cyg}$} category. The absorption component is wider, but velocity separation of components is mostly small. Flux ratio of emission and absorption has a large span. As one of the physical interpretations see \cite{2013MNRAS.431.2673K} for demonstration of the quasi-stationary appearance of the red-shifted absorption in T Tauri stars due to the unstable regime of the accretion.

Category {\bf S$_{abs}$} describes profiles with self--absorption. Emission is roughly centred on the nebular velocity and absorption can be blue or red shifted.

The last category {\bf E$_{abs}$} identifies profiles with emission in absorption, which may be normal stellar (photospheric) H$\alpha$ absorption. Narrow emission is moderately blue-shifted and the flux ratio has a large span where either component may be dominant.

\subsection{Temporal variability from repeated observations}

\begin{figure}[!htp]
   \centering
   \includegraphics[trim = 0 0 0 130mm, clip, width=\linewidth]{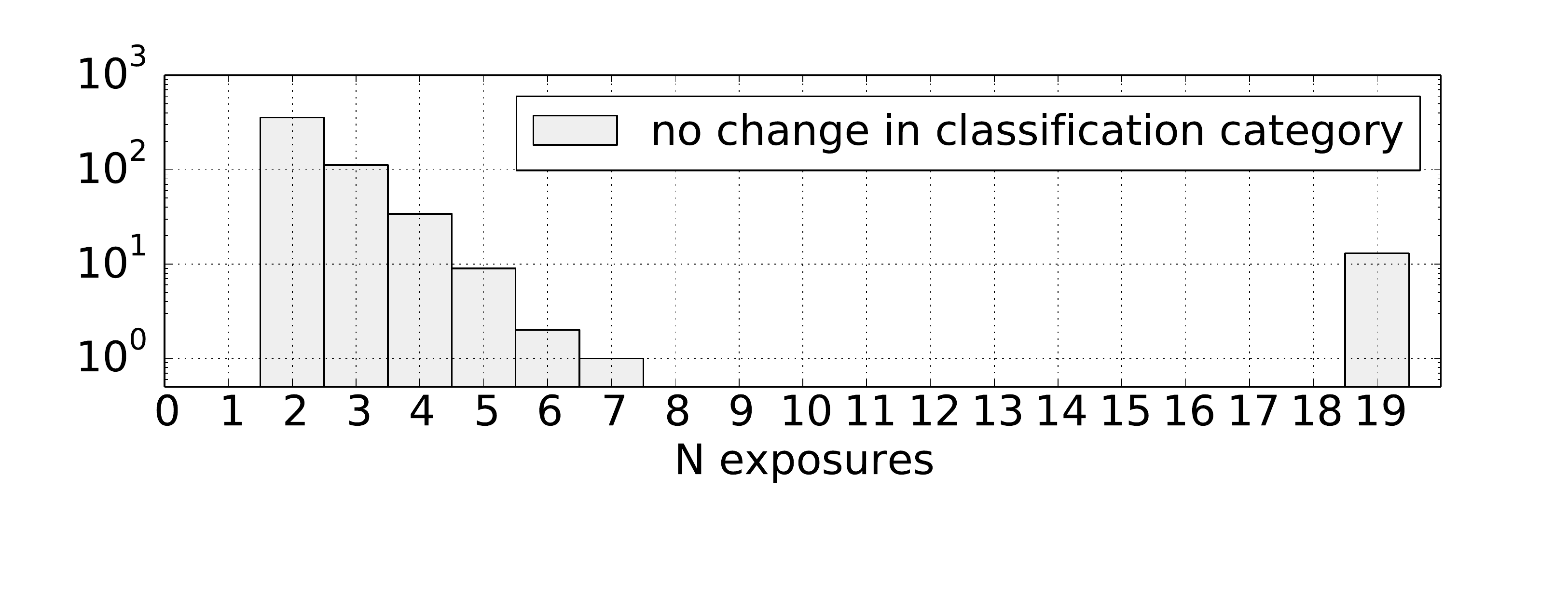}                        
   \caption{Objects with stable profile shape. Horizontal axis represents the number of observations (spectra) for each object and the number of objects is on the vertical axis. There are altogether 660 objects belonging to a unique classification category. Among them, 456 objects were observed twice, 134 sources were three spectra, and the remaining 70 sources have over 3 observations.}
   \label{temp_change}
\end{figure}

The temporal change of the profiles can be evaluated first in terms of classification categories from Table \ref{phys_cat}. We identify 660 out of 4459 objects that exhibit no change in classification category through time (Fig. \ref{temp_change}), and can be labelled as objects with stable profile shape.

\begin{figure}[!htp]
   \centering
   \includegraphics[width=0.49\textwidth]{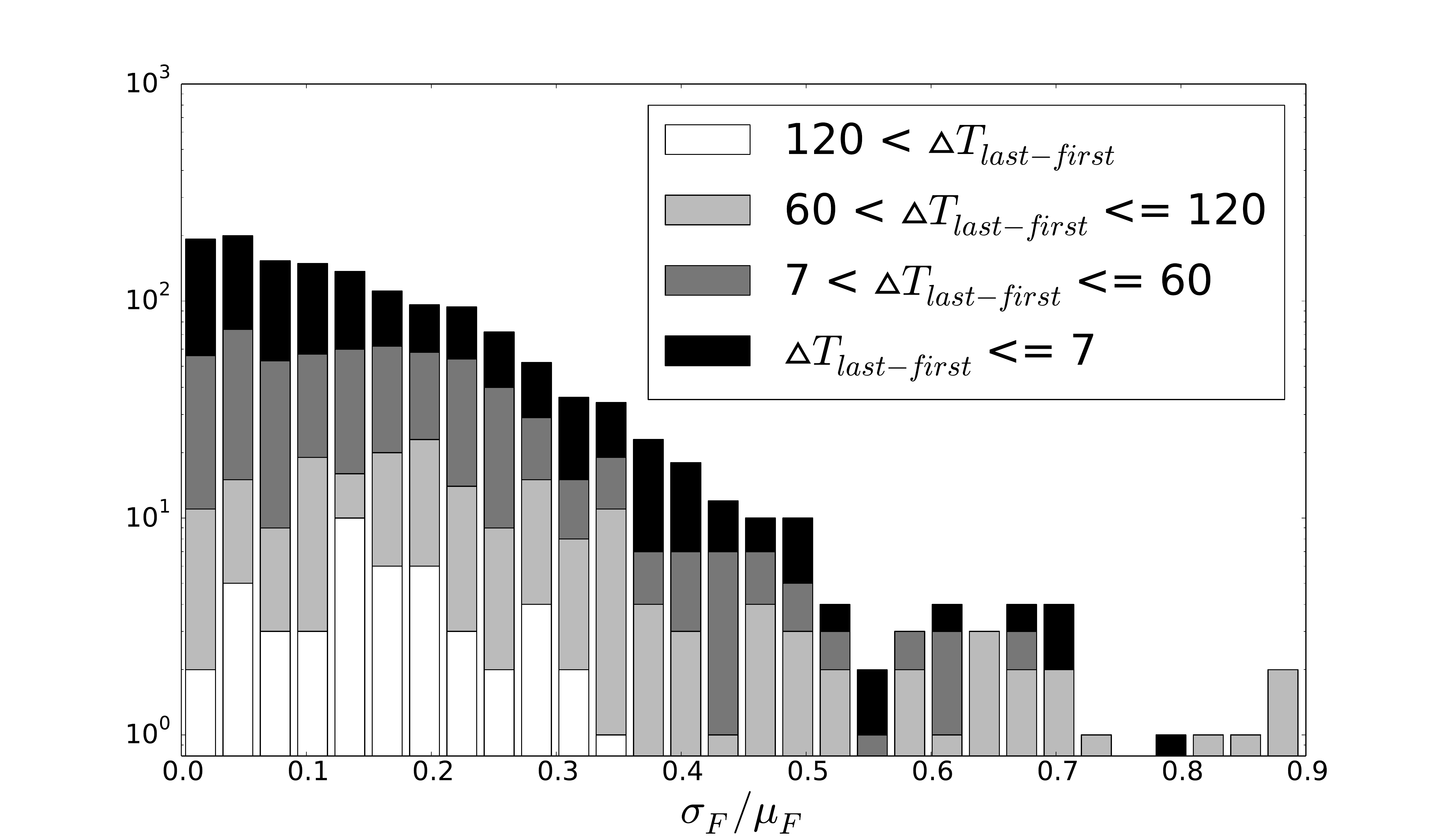}

   \caption{Distribution of variability $\sigma_F / \mu_F$ (standard deviation of $F$ divided by its mean) for the 1430 objects with more than one observation. The histogram shows four groups of different timespan ranges $\vartriangle T$ in days. Objects with the longest timespan (white) show small variability, while some of those with the shortest timespan (black) can exhibit significant changes. Objects within the group of the third longest timespan (light grey) display the strongest variation of their H$\alpha$ profile.}
   \label{varflux}
\end{figure}

Furthermore, we can measure the variability of H$\alpha$ profiles of a certain object by comparing their flux $F$ (Equation \eqref{integral}). By obtaining the standard deviation of the flux ($\sigma_F$) divided by its mean ($\mu_F$) for all spectra of a certain object, we can get an unbiased estimate of the flux variability whose distribution is shown in shown in Fig. \ref{varflux}. We note that no timespan between the first and last observations is longer than 1 year and we observe no evident correlation between this timespan and the flux variation.

The variability index displayed in Fig. \ref{varflux} is a measure of the profile change, where a high value of $\sigma_F / \mu_F$ almost certainly indicates a strong variability while a small value of the same parameter does not exclude relevant variations of the profile shape, because it is only related to the integrated flux. Nevertheless, we can assess which objects with repeated observations exhibit minimum temporal variability. This can be done by taking objects with 0 relative change of classification category (Fig. \ref{temp_change}) and taking the variability of flux $F$ from Fig. \ref{varflux} into account. A more complete variability analysis is out of scope of this work and will be addressed in future papers dealing with particular physical classes of stars with H$\alpha$ emission.

\subsection{Resemblance of categories}

According to our results, there are some cases of objects for which the classification of the H$\alpha$ profile changes with time. This might be due to the significant change in the actual profile shape or to the discrete conditions in our morphological classification scheme. The correlation of one category with the other is indicated by the off-diagonal elements in Table \ref{classbyclass}. Generally, there is one other category (column) per each category (row) that stands out and points to their correlation.

\begin{figure}[!htp]
   \centering
   
   \includegraphics[width=0.49\textwidth]{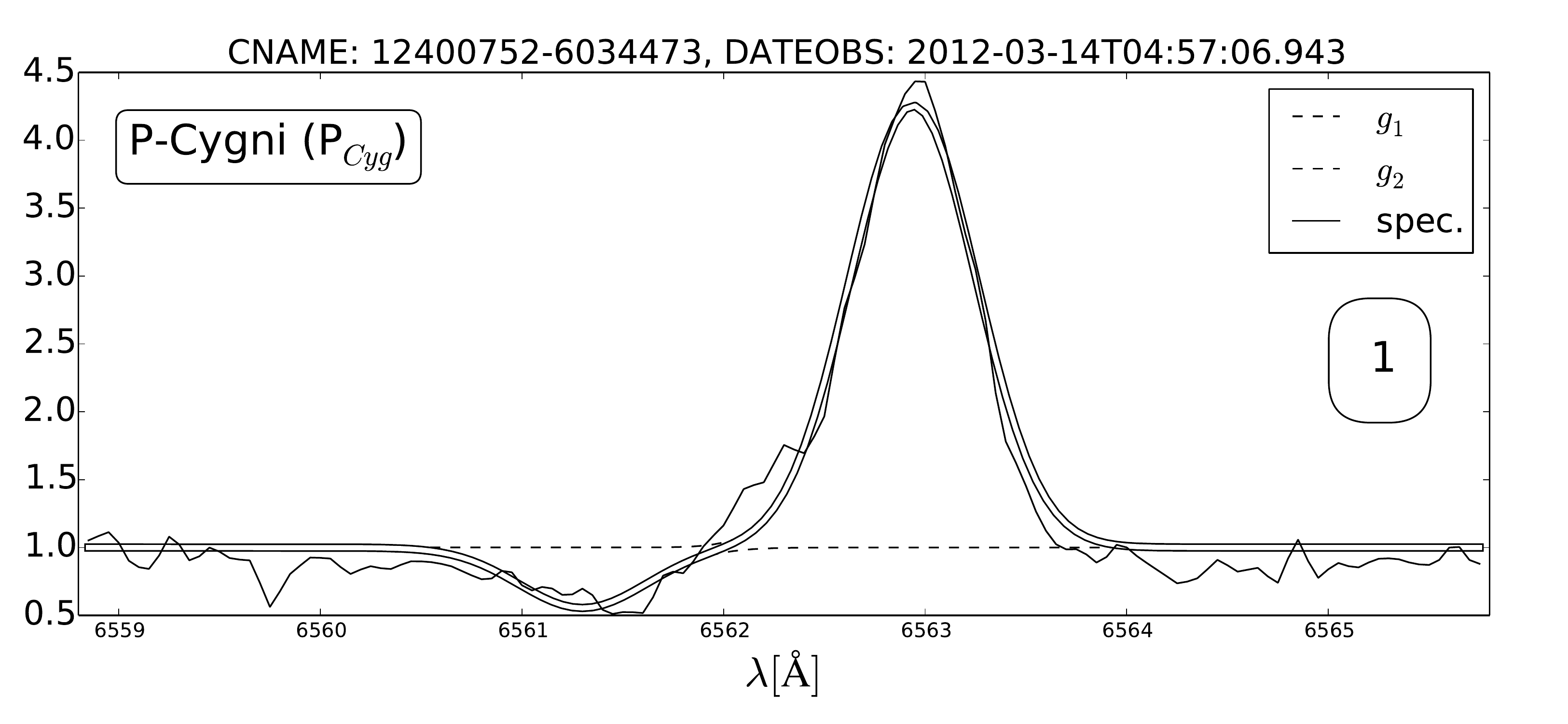}
    \includegraphics[width=0.49\textwidth]{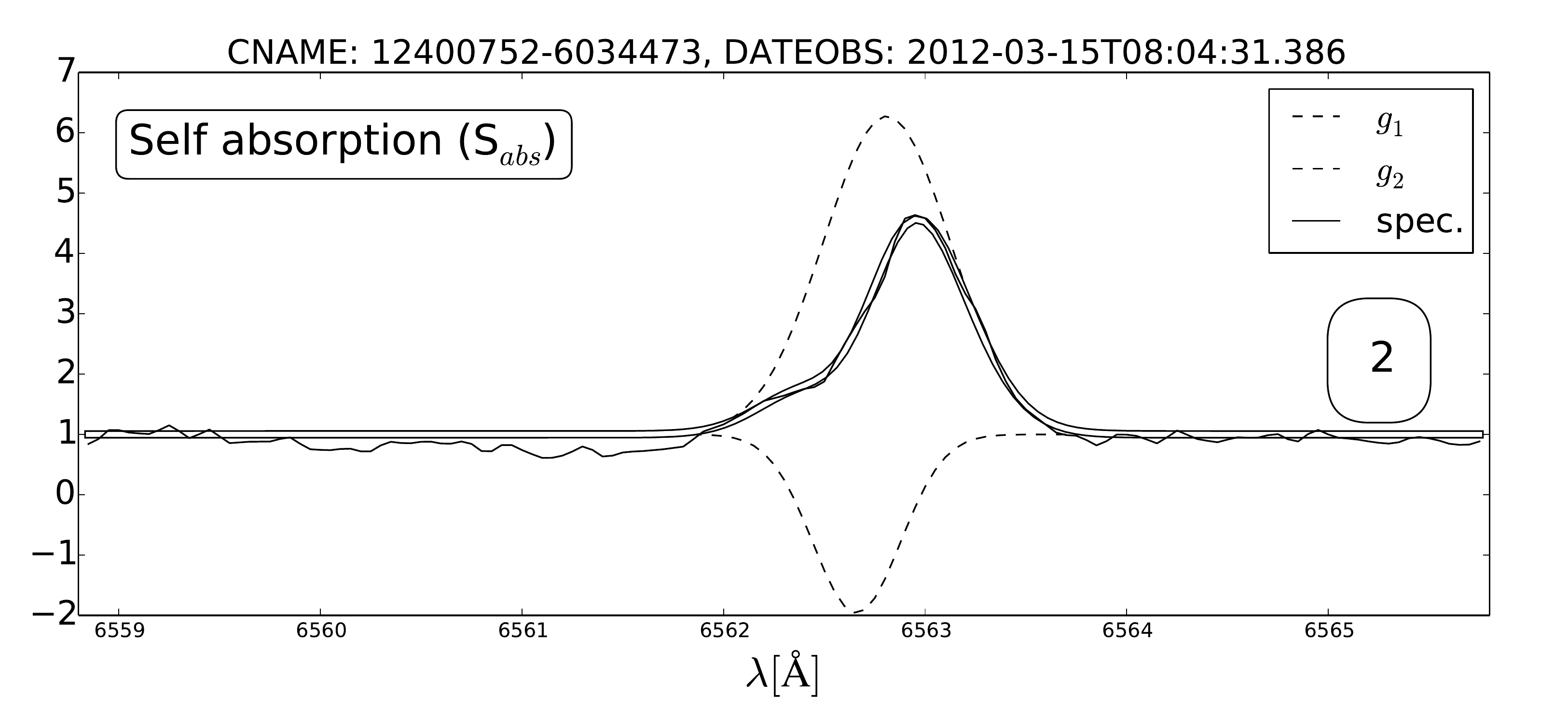}
    \includegraphics[width=0.49\textwidth]{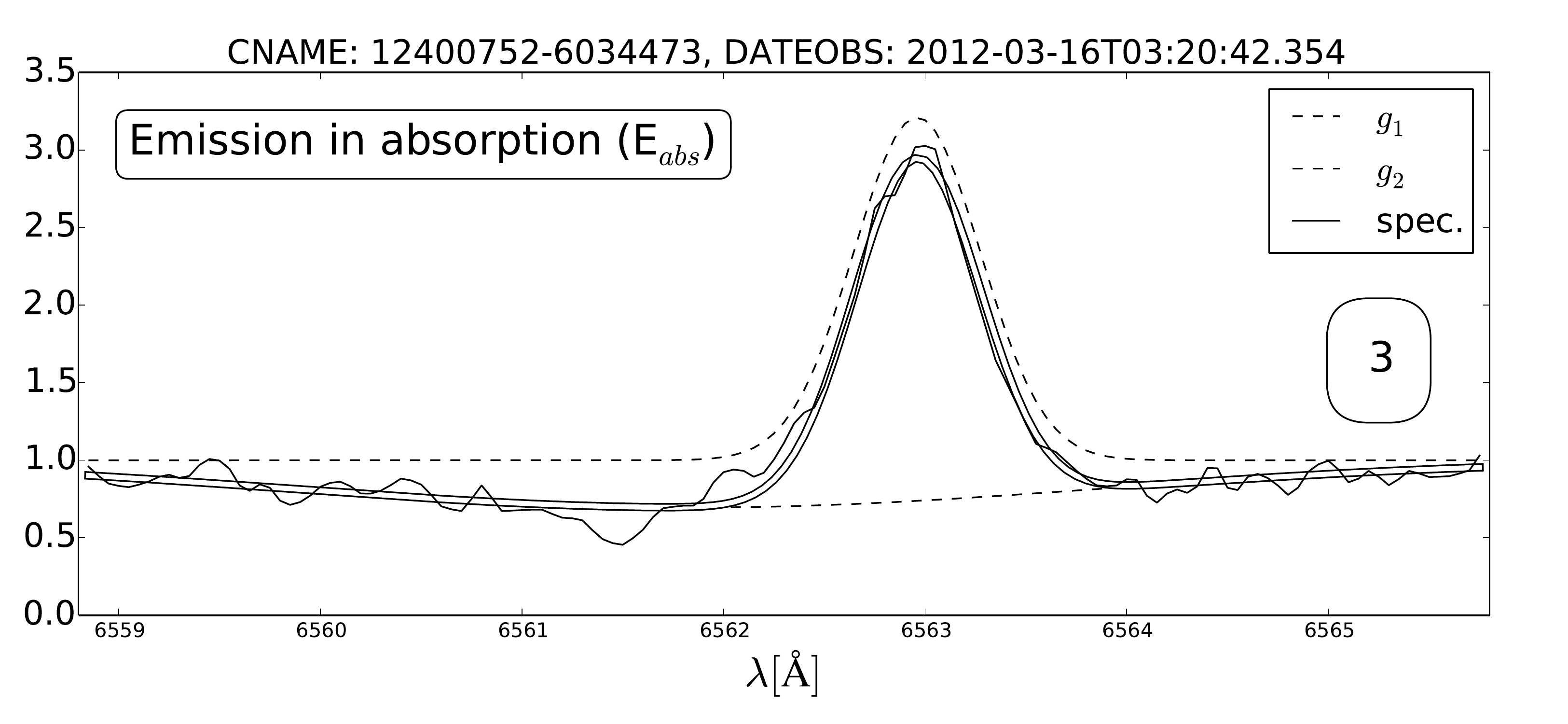}
    \includegraphics[width=0.49\textwidth]{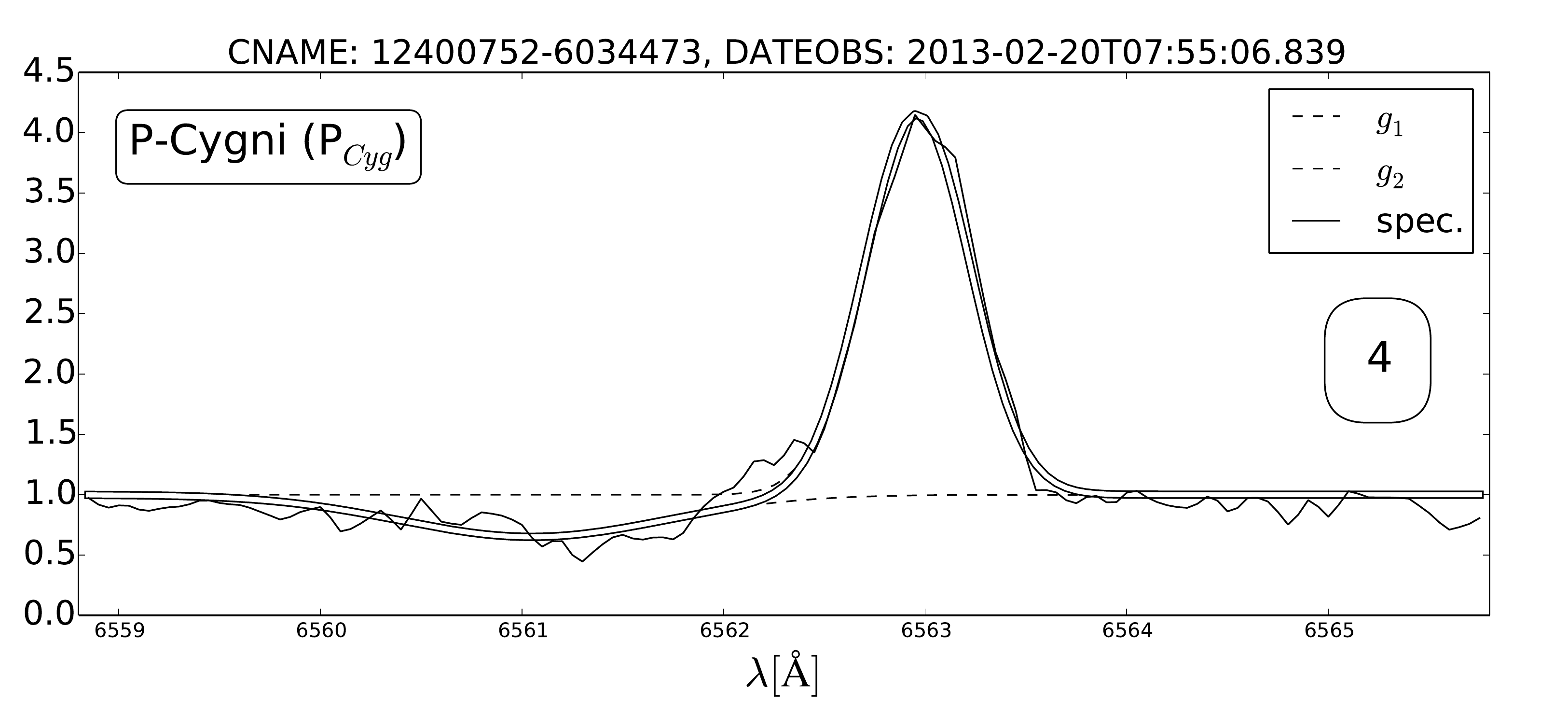}
    \includegraphics[width=0.49\textwidth]{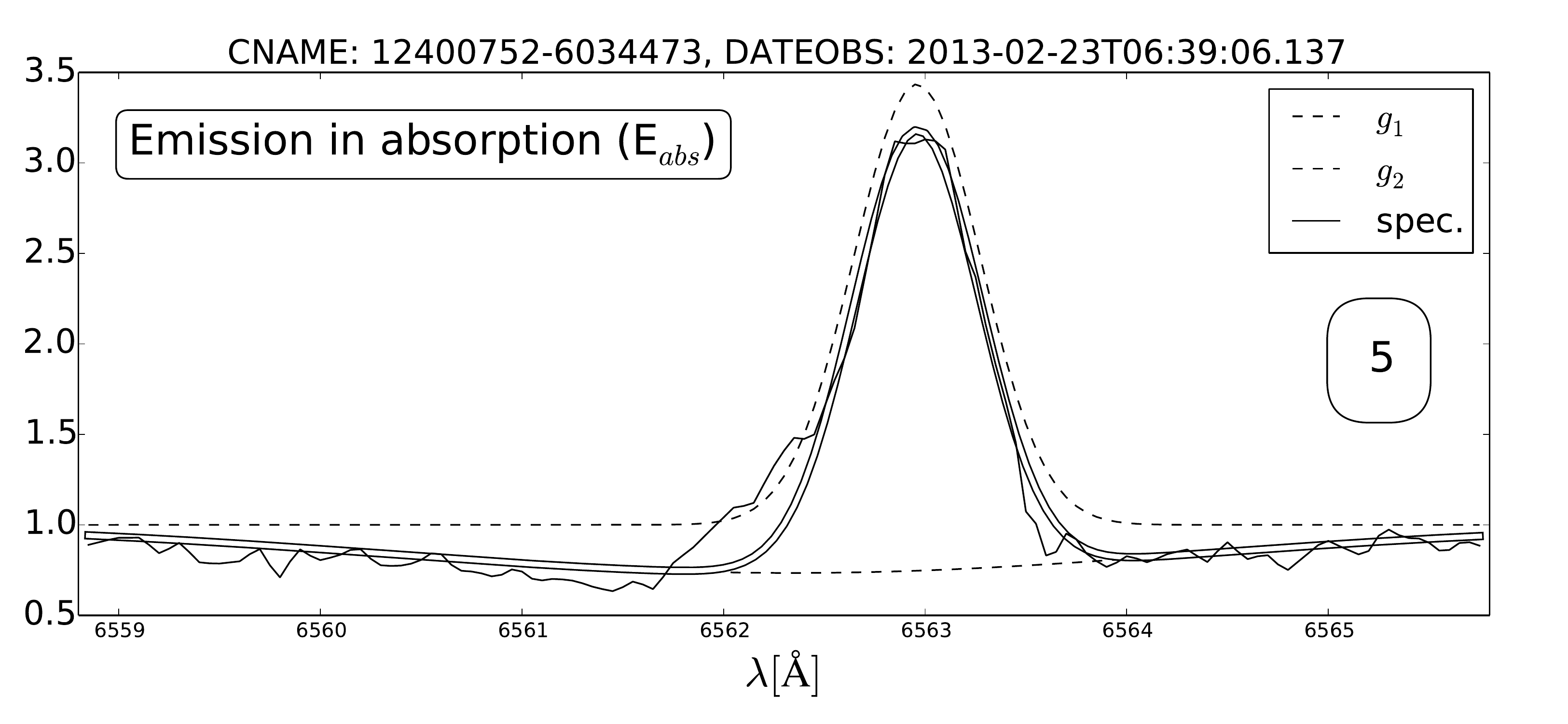}
    
   \caption{Typical spectrum of object with a pronounced intrinsic variability of emission
profiles for which the categories {\bf P$_{Cyg}$} and {\bf E$_{abs}$} are prevalent. Spectra are ordered by date of observation given in the title of each panel. The sum of all fitted components is represented by the thick black-white line. Continuum is at 1.}
   \label{pcyg}
\end{figure}

When {\bf E$_{bl}$} is the prevalent category for an object, it is most often in combination with {\bf S$_{abs}$}, which is best explained by one of the two components  being in transition between absorption and emission. Similarly, the category {\bf E$_{sp}$} can change to {\bf E$_{abs}$} if a relatively weak absorption is constantly present and one of the emission peaks diminishes. Alternatively, one of the emission peaks can be relatively weak, as indicated in Fig. \ref{sharp-peak} (panel (f)) and in Fig. \ref{pcyg} (panel 5), which could result in the preferential fitting of a wide, low-amplitude absorption in the best fit procedure.

The largest off-diagonal element connects the prevalent category {\bf E$_{dp}$} with {\bf S$_{abs}$}. The distinction between the two categories is largely influenced by the inclination of the slopes in the profile that are liable to change in the presence of additional weaker components or they are harder to retrieve in the case of more noisy spectra.  

The categories {\bf P$_{Cyg}$} and {\bf IP$_{Cyg}$} both exhibit noticeable connection to {\bf E$_{abs}$}. This can be due to their morphological similarity in cases where the fitted profile properties are close to the limiting conditions from Table \ref{phys_cat}, where it is sufficient that one of the fitted components' width or their separation changes slightly and the category shifts (e.g. due to the intrinsic change in velocity or orientation of the stellar accretion/outflow). Another option is demonstrated with the object in Fig. \ref{pcyg}, where the change of {\bf P$_{Cyg}$} to {\bf E$_{abs}$} category is due to intrinsically different profiles. The width of the absorption component in panels 1 and 4 is very different from that in panels 3 and 5. We also notice how the bluer side of the emission component changes, due to the presence of another emission component, and how this influences the morphologically different fit in panel 2. The object in Fig. \ref{pcyg} has a pronounced and intrinsic variability of its H$\alpha$ profiles. This is justly
reflected also in our morphological classification scheme. Such intrinsic changes are however relatively rare, as indicated by the prevalence of diagonal elements in Table \ref{classbyclass}.

\section{DISCUSSION AND CONCLUSIONS} \label{concl}

Gaussian fitting is one of the possibilities for analysis of single and multicomponent H$\alpha$ profiles. The methods that we discussed here were used to first detect and then classify spectral line profiles in a uniform and unbiased manner. In this paper we limit ourselves to the H$\alpha$ line, although this method could be applied to other spectral lines as well. From the 22\,035 GIRAFFE spectra in the GES internal data release, we found 9265 spectra displaying emission-like features. This amounts to 4566 out of 12\,392 unique objects, which represents a 36.8\% detection rate. The high percentage of detected emissions is expected since a large fraction of all objects observed by GES with the HR15N setup, which includes the H$\alpha$ line, are in the field of young open clusters. 

The Gaussian fitting produces results for 8869 out of 9265 analysed spectra, the difference being mainly due to the spectra with negative fluxes (see Appendix \ref{sec:detect}). Fitting is done with two independent Gaussian profiles and a third one to account for the nebular emission component, whenever necessary. The parameters of these three Gaussian profiles enable us to construct a simplified morphological classification. First, we flag all spectra (Appendix \ref{sec:identification}) following the ruleset in Fig. \ref{class_scheme} and summarise the result in Table \ref{title4}. Less than 0.3\% or 23 out of 9265 spectra are flagged as having only absorption components, which was subsequently confirmed by eye inspection and therefore proves the efficiency of our detection routine. These spectra are excluded from further consideration where we use the classification scheme described in Table \ref{phys_cat} to assign morphological categories to all spectra given in the online catalogue, whose contents are described in Tables \ref{title}, \ref{title2}, and \ref{title3}. Thus we have 8846 H$\alpha$ emission spectra in the catalogue corresponding to 4459 unique objects. The list of ADS references includes records about 305 or 7\% of these objects (Table \ref{title3}), while for roughly 25\% of them, the presence of H$\alpha$ emission is already indicated by SIMBAD.

We find that 2239 analysed objects (4049 spectra) have an indication of the presence of a nebular emission. This evaluation depends on the assumption that nebular [N~II] emission lines correlate well with H$\alpha$ emission line and that the nebular emission is more or less smooth over a certain field of view or across individual clusters, so that the width and radial velocity of nebular lines do not show abrupt spatial changes. Therefore, $\lambda_{n}$ and $\sigma_{n}$ parameters were fixed during the Gaussian fitting procedure. Although the [N~II] emission lines can be explained by the nebular contribution, we caution that they can also be present due to the outflow activity of the star, so it is possible that in some cases, the fit of the nebular component corresponds to some other than the strictly nebular contribution to the line profile. 

Overall statistics of the fitted components to H$\alpha$ profiles is given in Fig. \ref{params}. We note that the high values of $A$ parameter mostly correspond to the very strong emission lines while sometimes, the best fit was produced with two strong Gaussian components, one in emission and the other in absorption, that summed together to represent a relatively weaker component. The nebular (sky) lines are fitted with only one Gaussian profile to avoid degeneracy issues, although in some cases, we acknowledge the possible existence of several prominent sources of sky emission (e.g. in the line of sight toward Trumpler 20 and NGC 4815 open clusters), that can effectively broaden the fitted nebular component ($\sigma_n$ values up to limiting 0.5 \AA). However, in such cases, the nebular component is usually excluded from the fit (lack of $\sigma_n$ values above 0.4 {\AA} in Fig. \ref{params}) due to the conditions listed in Appendix \ref{gauss_fitting}. 

Through an unbiased way we demonstrate that 3765 objects (7698 spectra) exhibit intrinsic H$\alpha$ emission, which is formed in the immediate vicinity of the observed object. This accounts for 30\% of all 12392 distinct objects investigated in the present study. The flags from Table \ref{title4} and conditions in Table \ref{phys_cat} further enable us to present the different morphologies of observed H$\alpha$ profiles in the context of categories, which aim to identify and describe physically distinct groups of objects. The classification categories are presented in Figs. \ref{Em_blend}-\ref{Em_in_Abs} where the parameters of fitted Gaussian profiles ($g_1$, $g_2$) and the typical profiles illustrate the contrast between them. 

The selection and definition of classification categories are based on previous experience with the different types of emission line profiles that are commonly observed in stellar spectra. Such profiles already have diverse physical interpretations and by assigning categories to our objects we merely aim to guide towards them. We note that the conditions in Table \ref{phys_cat} are discrete, which can cause arbitrary transitions between categories for spectra of individual objects. This can also be tested by the reader with the use of the first five best fit solutions given in Table \ref{title}. We also note that the order and the timespan of the consecutive changes of categories through time is not accounted for in Table \ref{classbyclass}. We therefore caution that sometimes, rapid changes of categories for individual objects can occur, even when one category is prevalent. We can additionally comment on the choice of one of the parameters from Table \ref{phys_cat} by examining how the classification behaves when we fix $K_{EE} = K_{Cyg} = 0.9$ and vary only $K_{sp}$. It can be shown that for our dataset, the first stability region or constant values of affected diagonal elements from Fig. \ref{classbyclass} arise just at $K_{sp} \approx 50$ km s$^{-1}$. 

Repeated observations for 1430 objects enable us to derive some basic properties about their temporal variability as presented in Figs. \ref{temp_change} and \ref{varflux}. Furthermore, 1191 objects with multiple spectra having two intrinsic components help determine the values of parameters $K_{EE}, K_{Cyg}$, and $K_{sp}$ for our morphological classification scheme. Spectra in Fig. \ref{pcyg} demonstrate the change of profiles with time and at the same time point to some of the characteristics of our classification, which enables us to select a particular type of profile, caused by a specific underlying physics, and helps better understand the corresponding stellar family. 

A more detailed investigation of variability could extend to more emphasised study of line profile intensities, continuum variability, analysis of broad band photometry and SEDs for our sources, which can be obtained with the help of VOSA -- Virtual Observatory SED Analyzer \citep{2008A&A...492..277B}. Such subsequent analysis is beyond the scope of this work but might be pursued by other authors from the GES. Here, we present a more general approach with fewer fitted components to line profiles in order to obtain sensible H$\alpha$ classification categories for each spectrum. It becomes harder to reach this objective in a more detailed and model dependent approach. The methods used for analysing H$\alpha$ profiles and the results presented here could therefore form a valuable basis for future work.\\

\bibliographystyle{apj}
\bibliography{sample}

\clearpage
\begin{appendices}

\section{DETECTION AND GAUSSIAN FITTING OF H$\alpha$ EMISSION LINES} \label{sec:detect}

When classifying H$\alpha$ features we aim at separating line profiles exhibiting emission components from those with absorption-only profiles. We start with selecting 22\,035 object spectra before background sky subtraction and continuum normalisation. This is important since sky subtraction can introduce spurious effects if the sky (nebular) H$\alpha$ component is not correctly identified and treated for each individual spectrum. Due to the large volume of our sample, the most practical way to attempt the detection of emission lines is by means of an automated procedure that we describe below. Gaussian fitting and its details are described next.

\subsection{Detection} \label{subdet}

The method employed for detection and parametrisation of H$\alpha$ line profiles includes several steps and conditions. The wavelength values used throughout this work are heliocentric. Normalisation of spectra is done first by extracting only the wavelength ranging from 6549.8 to 6582.8 {\AA} so as to avoid the [N~II] emission lines and then calculating the median of the outermost 10 {\AA} on each side. Division of spectra with this median value then yields a satisfactory normalisation, as demonstrated in Fig. \ref{spectra}.

The adopted central wavelength of H$\alpha$ is $\lambda_0 = 6562.83$ {\AA}. We often notice strong emission lines at $\sim 6553$ {\AA} and $\sim 6577$ {\AA} in our sample of spectra and so we define the left (6555 {\AA}) and right (6575 {\AA}) boundary in order to avoid false detections when searching for H$\alpha$ emission. We note that the mentioned emission lines may be of atmospheric origin due to [OH] excitation \citep{1997PASP..109..614O}. The selected region of spectra between the initial boundaries defined above is in our experience large enough to include the whole H$\alpha$ profile. Next, we set the first condition for exclusion of H$\alpha$ absorption: 
\begin{equation} \label{pogoj1}
1 - min({\bf y}) > 3 (max({\bf y}) - 1)
\end{equation}
where the value 1 denotes the continuum level, and ${\bf y}$ is the flux of the selected normalised region. If this condition is satisfied, the spectrum is not considered any further, which means that we can only detect emissions if the amplitude of the highest peak above the continuum in the selected normalised region is at least one third of the amplitude of the strongest absorption. Note that we therefore refrain from detection of faint chromospheric emission activity, which sits on a strong absorption component. Treatment of such weak emission is beyond the scope of this paper because it would require a detailed modelling of the stellar atmosphere. We refer the readers to \citet{lanzafame_corrected} and \citet{A.Frasca_corrected} for a discussion of this topic. If the above condition is not fulfilled, the procedure continues by automatically inspecting the values of ${\bf y}$ by going step by step (pixel by pixel) from the left initial boundary (6555 {\AA}) to the right (6575 {\AA}) and vice-versa. The main task here is to define the smallest region in which emission is located and to exclude false detections due to strong noise or cosmic rays. We therefore need to accurately define the limits of the potential H$\alpha$ emission as near to the observed profile as possible. It turns out it is better to set two conditions to achieve this goal, first for broader and second for narrower lines (narrow emission over wide absorption): 
\begin{footnotesize}
\begin{align}
{\bf y}[k + 40] - {\bf y}[k] > d \quad &\& \quad {\bf y}[k] \geqslant 1 \label{pogoj2}\\
{\bf y}[k + 10] - {\bf y}[k] > 2.5 d \quad &\& \quad 1 - {\bf y}[k] < (max({\bf y}) - 1) / 3 \label{pogoj3}
\end{align}
\end{footnotesize}
where $k$ is the running pixel index and $d = (max({\bf y}) - 1) / 5$. One pixel corresponds to $0.05$ {\AA}. These two conditions are valid for left to right direction, and the ones for reverse are  symmetrical. The second argument of both conditions restricts the value of flux to either above 1 (continuum) for wider emission or below 1 for relatively narrow emission over wider absorption. When the two conditions fail, the spectra are excluded from further consideration. 

Although they might seem arbitrary, the conditions \eqref{pogoj1}, \eqref{pogoj2}, and \eqref{pogoj3} are set according to our experience for the best overall detection performance. 11\,230 spectra out of 22\,035 are selected when making use of the condition \eqref{pogoj1}, while 1965 are subsequently excluded with conditions \eqref{pogoj2} and \eqref{pogoj3}, among which there are also some realistic emission profiles. We therefore note that the described procedure is not perfect and its performance also depends on the S/N. Besides excluding emission profiles, the above conditions are also liable to produce false emission detections in case of strong noise or cosmic rays in the spectra. Generally, such cases are few and an automatic procedure usually allows us an accurate and rapid detection of H$\alpha$ emission for thousands of spectra. Out of 22\,035 spectra, 9265 are identified as emission-type profiles. Many of these spectra are those of objects with multiple observations both in the GES internal data release or ESO archive. Therefore there are 4566 unique objects in our sample. 

\subsection{Gaussian fitting} \label{gauss_fitting}

We perform more thorough analysis of our selected sample of 4566 objects and their 9265 spectra by means of Gaussian fitting of the spectral region around H$\alpha$. This is also where we tackle with the problem of identification of the nebular component, which might contribute an important part of the emission flux if not being the only source of emission. As already mentioned, the majority of targets in our sample are stars in young clusters and star forming regions, which are often wrapped in a nebular environment, so it is essential we separate this external contribution of flux in H$\alpha$ from the one intrinsic to a given object.

All Gaussian fitting referred to in this paper is done with the least-squares Levenberg-Marquardt algorithm. We use a sum of two independent Gaussian components $g_1, g_2$ and one optional $g_n$ to account for the nebular component. The nebular component of H$\alpha$ is indicated by the presence of [N~II] forbidden lines (6548.05 {\AA}, 6583.45 {\AA}) in the spectra as well as by the sky spectra in the vicinity of observed targets. Whether we include $g_n$ into the fitting routine depends on the assessment of the two mentioned indicators.

The fitting procedure starts by putting the continuum of normalised spectra (see Sect. \ref{subdet}) to 0. We disregard the 396 spectra with low S/N ($\sim 4$) displaying negative fluxes.

Next, we evaluate the indicators for the presence of the nebular H$\alpha$ component. The preferred indicator is the presence of [N~II] emission around H$\alpha$. Both [N~II] forbidden lines are fitted with Gaussian profiles ($g_{N1}, g_{N2}$) with Gaussian parameters ($\lambda_{N1}, \sigma_{N1}, A_{N1}, \lambda_{N2}, \sigma_{N2}, A_{N2}$) where $\lambda_{N1}$ and $\lambda_{N2}$ are allowed to shift $\pm 1$ {\AA} and the fitting range is within $\pm 2$ {\AA} of the canonical centre of the [N~II] lines. The $\sigma_{N1}$ and $\sigma_{N2}$ values fall in the range of 0.1 to 0.5 {\AA} and all three Gaussian parameters are then determined with the least-squares method. The three conditions that must be fulfilled to accept this first indicator are: (i) both $A_{N1}$ and $A_{N2}$ must be greater than 0.5 to compensate for noisy spectra, (ii) $\chi^2/pixel$ divided by $A_{N1,N2}$ must be $< 1.0$, and (iii) $| |\lambda_{N1} - \lambda_{N2}| - (6583.45$ {\AA} $ - 6548.05 $ {\AA} $) | < 0.1$ {\AA}. If these 3 conditions are met, the wavelength parameter $\lambda_{n}$ for $g_n$ is fixed to $\lambda_{n} = \lambda_{N1} + (6562.83 - 6548.05)$ {\AA} and the $\sigma_{n}$ parameter is fixed to the average of $\sigma_{N1}$ and $\sigma_{N2}$ multiplied by the ratio H$\alpha_{\sigma}$/NII$_{\sigma}$ explained in the next paragraph. When these conditions are not satisfied, we continue with evaluation of emission in sky spectra.

Out of 132 available fibres for Giraffe spectrograph, around 24\% in the full sample of 28\,957 spectra are used to record background sky spectra and these provide valuable information about the object's surroundings. Ideally, there should be one sky (background) exposure per object, which would enable us to more accurately determine the nebular contribution to the flux from stars in young clusters. Here we take five spectra from nearest surrounding fibres closest to each of the observed objects. Maximum distance is set to $5^{\circ}$ but usually all five fibres are located within $3'$ of the object fibre, which roughly translates to 1 pc at the distance of 1 kpc. These five sky spectra are averaged and all three emission lines (H$\alpha$ and both [N~II]s) are fitted in the same manner, as described in the previous paragraph. The ratio H$\alpha_{\sigma}$/NII$_{\sigma}$ = 1.3 is obtained from average values of $\sigma_{N1}, \sigma_{N2}$ and $\sigma_{n}$ in all evaluated sky spectra. Similarly to the first indicator of nebular emission, the same three conditions are applied here together with an additional one stating that $| \lambda_{n} - 6562.83$ {\AA} $ | < 0.1$ {\AA}. 

When neither of the conditions from the previous two paragraphs are satisfied, $g_n$ is not included in the fitting procedure, hence we only use $g_1$ and $g_2$. The source/lack of nebular indication is noted in Table \ref{title}. The $g_n$ component has $\lambda_{n}$ and $\sigma_{n}$ fixed and only $A_{n}$ is fitted. For $g_1$ and $g_2$ the $\lambda_{1,2}$ and $\sigma_{1,2}$ parameters vary in a certain interval with a fixed step whereas $A_{1,2}$ is also fitted by the fitting procedure. The step size of $\sigma_{1,2}$ is the same for all spectra. We set the lower value to 0.15 {\AA} based on the resolution of the spectrograph (R $\sim$ 17\,000) and the upper value to 2 {\AA} that is sufficient for the broader components of the profiles. The use of 15 values that are logarithmically spaced allows us to better cover the regime of narrow lines to the detriment of broader ones. The $\lambda_{1,2}$ parameter has 21 linear intermediate values, which cover the range between the lower and upper boundaries, but the boundaries vary from spectrum to spectrum as described in the following paragraph.

We aim to fit the smallest possible range including the whole H$\alpha$ profile. This is partly done in the emission detection stage described at the beginning of this section, but here we want to define the region with absorption as well as emission parts of a generally multicomponent H$\alpha$ profile. We therefore limit the fitting range first to 5 {\AA} ($\sim 230$ km s$^{-1}$) at each side of $\lambda_0$ to avoid the already mentioned emission features and to ensure that the maximum initial grid step for $\lambda_{1,2}$ parameters in the fitting procedure does not exceed $0.5$ {\AA}. Next, the procedure proceeds from the middle to each side pixel by pixel (1 pixel is 0.05 {\AA}), evaluating each consecutive chunk of 20 pixels ahead for the change in sign of the residual spectrum (normalised flux -- 1). When the change in sign occurs at least three times in a given interval of 20 pixels, the procedure assumes that the continuum is reached and sets the furthermost pixel of the current chunk to be the limiting boundary. If this condition is not reached, the boundary (the limiting interval [$x_{min}, x_{max}$]) stays at $\pm 5$ {\AA} from $\lambda_0$. We use 21 linearly spaced values to split the interval between the lower and upper boundaries. From the intervals of values for $\lambda_{1,2}$ and $\sigma_{1,2}$ we get 49\,770 combinations of $g_1$ and $g_2$ that can be summed together with $g_n$ according to criteria in the above paragraphs.

The results of the fit give us the values of $\lambda_{1,2,n}, \sigma_{1,2,n}$ and $A_{1,2,n}$ of the best fitted $g_1 + g_2 + g_n$, along with the reduced $\chi^2$ ($\chi^2$ per pixel). The same procedure is then repeated once more assuming $\lambda_1 = \lambda_2$, as already described in Sect. \ref{sec:data}, and the final adopted set of parameters is the one achieving a better match between observed and modelled H$\alpha$ profiles.

\subsection{Fit statistics} \label{gauss_fitting_results}

\begin{figure*}[!htp]
   \centering
   \includegraphics[trim = 25mm 0 25mm 0, clip, width=\textwidth]{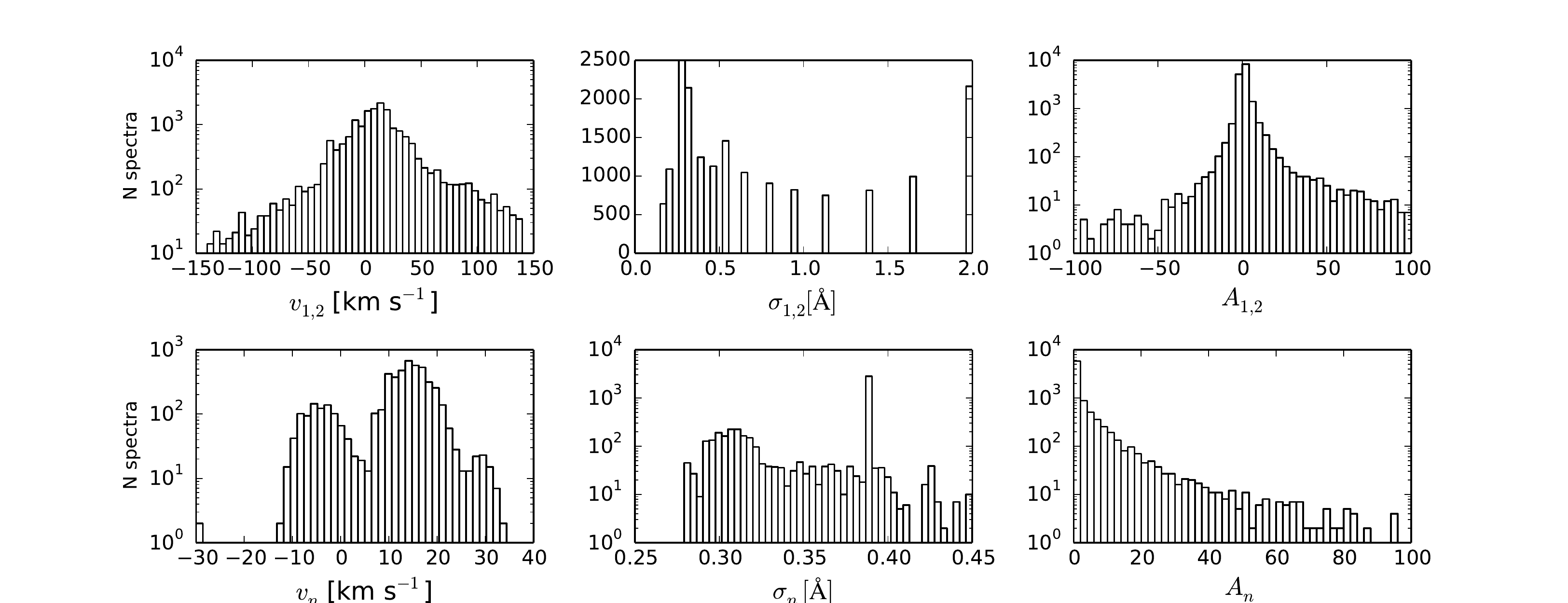}                        
   
   \caption{Histograms of Gaussian parameters for the 4459 emission type objects. We show the values for components $g_1$ and $g_2$ (top panels), along with those for $g_n$ (bottom panels). When the nebular component is not included in the fit, their values are left out, except for $A_n$ that are stacked in the first bin (value = 0).
   }
   \label{params}
\end{figure*}

 The distribution of the values obtained for the parameters describing $g_1, g_2$, and $g_n$ are shown in Fig. \ref{params}. The double-peaked profile in the $v_n$ histogram implies that objects with indicated nebular component belong to distinct nebular environments or clusters with different radial velocity. Histogram of $\sigma_n$ shows a strong peak at 0.39 {\AA}, which corresponds to the instrument resolution, indicating a nebular emission with an intrinsic FWHM smaller than the instrumental profile. 

The $\chi^2_r$ parameter might be misleading where one or more emission components are relatively strong, therefore we construct another measure of the goodness of fit - $\chi^2_{rf}$, which is the previously defined $\chi^2_r$ normalised with average flux where the total flux of the line is the integral 
\begin{equation} \label{integral}
F = \int \left| g_1(\lambda) + g_2(\lambda) + g_n(\lambda)\right| \mathrm{d}\lambda 
\end{equation}
over the whole fitting range ([$x_{min}, x_{max}$], see above) determined for each spectrum. The distributions of $\chi^2_r$ and $\chi^2_{rf}$ for all spectra are plotted in Fig. \ref{chi2}. Due to the normalisation of $\chi^2_r$ for the strongest emissions, the tail on the right is reduced and the distribution shifted to lower values when going from $\chi^2_r$ to $\chi^2_{rf}$. The bump on the left of the $\chi^2_{rf}$ distribution is mainly due to the very strong nebular emissions where $\chi^2_r$ is small and $F$ is relatively large.

\begin{figure}[!htp]
   \centering
   \includegraphics[width=\linewidth]{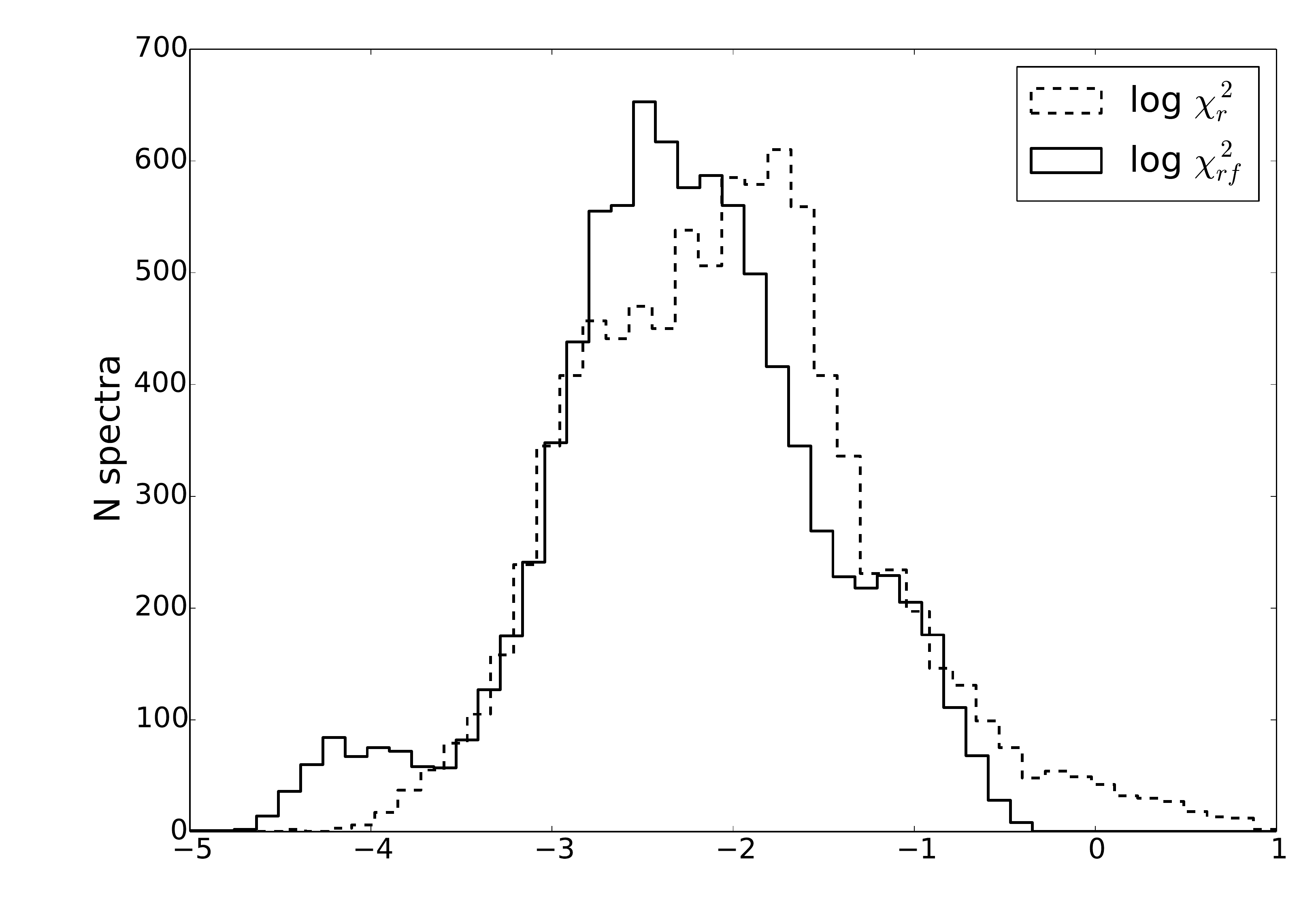}                        
   \caption{Histograms of $\chi^2_{rf}$ (solid line) and $\chi^2_r$ (dashed line) representing the formal goodness of fit for 8846 spectra. The horizontal axis is logarithm of $\chi^2_r$ and $\chi^2_{rf}$, where square root of $\chi^2_{rf}$ represents relative O-C.}
   \label{chi2}
\end{figure}

\section{IDENTIFICATION OF FITTED COMPONENTS} \label{sec:identification}

In order to implement our morphological classification of H$\alpha$ line profiles, we use the parameters of fitted $g_1, g_2$, and $g_n$ to determine their type, relative strengths and to what degree one blends with another. We define a set of rules, which enables us to assign one of the following flags to each spectrum: E, EE, EA, AE, N, EN, EEN, EAN, AEN. Letters E, A, and N denote intrinsic emission, intrinsic absorption, and nebular emission, respectively. The number of letters in each flag matches the number of components, with the exception of the least interesting ones for our study (i.e. N, AN, and AAN) where the emission is solely due to the nebular component. These three flags are therefore joined into flag N. The order of letters E and A represents the relative strength (absolute equivalent width) of the corresponding components where the first one is stronger (e.g. EA - absorption is weaker than emission, EAN - the same with additional nebular emission component). 

The rules are defined in the outline with conditional statements in Fig. \ref{class_scheme}, where one of the branches deals with flags without nebular emission and the other with those including nebular emission (N).

\begin{figure*}[!htp] 

{\centering
\underline{$f_{n} \leqslant \varepsilon_f$}\\
}

\begin{itemize}
   \small
          
    \item {\bf E:}
    $
        f_1 > 0 \quad\&\quad f_2 > 0 \quad\&\quad |\lambda_1 - \lambda_2| < \varepsilon_{\lambda} \quad\&\quad 
        |\sigma_1 - \sigma_2| < \varepsilon_{\sigma} 
        \quad||\quad
        |f_1| < \varepsilon_f \quad\&\quad f_2 > 0 
        \quad||\quad 
        |f_2| < \varepsilon_f \quad\&\quad f_1 > 0           
    $
    \item {\bf EA:}           
      $
      f_1 > 0 \quad\&\quad f_2 < 0 \quad\&\quad |f_1| > |f_2|
      \quad||\quad
      f_1 < 0 \quad\&\quad f_2 > 0 \quad\&\quad |f_2| > |f_1| 
     $
    
    \item {\bf AE:}  
      $ 
        f_1 > 0 \quad\&\quad f_2 < 0 \quad\&\quad |f_1| < |f_2|
        \quad||\quad
        f_1 < 0 \quad\&\quad f_2 > 0 \quad\&\quad |f_2| < |f_1| \quad\quad\quad
      $
    \item {\bf EE:}
    		$
      f_1 > 0 \quad\&\quad f_2 > 0 \quad\&\quad |f_1| \geqslant \varepsilon_f \quad\&\quad |f_2| \geqslant \varepsilon_f
      	$
\end{itemize}
\bigskip

{\centering
\underline{$f_{n} > \varepsilon_f$}\\
}
\begin{itemize}
  \small
    \item {\bf N:}              
      \begin{align*}  
        &|f_1| < \varepsilon_f \quad\&\quad |f_2| < \varepsilon_f 
        \quad||\quad\\      
        &|\lambda_1 - \lambda_n| < \varepsilon_{\lambda} \quad\&\quad |\lambda_2 - \lambda_n| < \varepsilon_{\lambda} \quad\&\quad |\sigma_1 - \sigma_n| < \varepsilon_{\sigma} \quad\&\quad |\sigma_2 - \sigma_n| < \varepsilon_{\sigma} \quad\&\quad f_1 > 0 \quad\&\quad f_2 > 0         
        \quad||\quad\\
        &|f_2| < \varepsilon_f \quad\&\quad |\lambda_1 - \lambda_n| < \varepsilon_{\lambda} \quad\&\quad |\sigma_1 - \sigma_n| < \varepsilon_{\sigma} \quad\&\quad f_1 > 0         
        \quad||\quad\\
        &|f_1| < \varepsilon_f \quad\&\quad |\lambda_2 - \lambda_n| < \varepsilon_{\lambda} \quad\&\quad |\sigma_2 - \sigma_n| < \varepsilon_{\sigma} \quad\&\quad f_2 > 0 \quad\quad\quad         
      \end{align*}   

    \item {{\bf EN} ({\bf AN} if the last conditional term in each line does not hold):} 
      \begin{align*} 
        &|\lambda_1 - \lambda_n| < \varepsilon_{\lambda} \quad\&\quad |\sigma_1 - \sigma_n| < \varepsilon_{\sigma} \quad\&\quad f_1 \geqslant 0 \quad\&\quad |\lambda_2 - \lambda_n| \geqslant \varepsilon_{\lambda} \quad\&\quad |f_2| \geqslant \varepsilon_f \quad\&\quad f_2 > 0
        \quad||\quad\\
        &|\lambda_2 - \lambda_n| < \varepsilon_{\lambda} \quad\&\quad |\sigma_2 - \sigma_n| < \varepsilon_{\sigma} \quad\&\quad f_2 \geqslant 0 \quad\&\quad |\lambda_1 - \lambda_n| \geqslant \varepsilon_{\lambda} \quad\&\quad |f_1| \geqslant \varepsilon_f \quad\&\quad f_1 > 0
        \quad||\quad\\
        &|\lambda_1 - \lambda_n| \geqslant \varepsilon_{\lambda} \quad\&\quad |f_1| \geqslant \varepsilon_f \quad\&\quad |f_2| < \varepsilon_f \quad\&\quad f_1 > 0
        \quad||\quad\\
        &|\lambda_2 - \lambda_n| \geqslant \varepsilon_{\lambda} \quad\&\quad |f_2| \geqslant \varepsilon_f \quad\&\quad |f_1| < \varepsilon_f \quad\&\quad f_2 > 0
        \quad||\quad\\
        &|\lambda_1 - \lambda_n| < \varepsilon_{\lambda} \quad\&\quad |\sigma_1 - \sigma_n| \geqslant \varepsilon_{\sigma} \quad\&\quad |f_2| < \varepsilon_f \quad\&\quad f_1 > 0
        \quad||\quad\\
        &|\lambda_2 - \lambda_n| < \varepsilon_{\lambda} \quad\&\quad |\sigma_2 - \sigma_n| \geqslant \varepsilon_{\sigma} \quad\&\quad |f_1| < \varepsilon_f \quad\&\quad f_2 > 0
        \quad||\quad\\
        &|\lambda_1 - \lambda_2| < \varepsilon_{\lambda} \quad\&\quad |\sigma_1 - \sigma_2| < \varepsilon_{\sigma} \quad\&\quad f_2 + f_1 > 0 \quad\quad\quad
      \end{align*} 

    \item {\bf AAN:}  
      $f_1 \leqslant 0 \quad\&\quad f_2 \leqslant 0$
    \item {\bf AN:}  
      $(f_1 \leqslant -\varepsilon_f \quad\&\quad |f_2| < \varepsilon_f) \quad||\quad (f_2 \leqslant -\varepsilon_f \quad\&\quad |f_1| < \varepsilon_f)$

    \item {\bf EAN, AEN, EEN:} The same conditions apply as for {\bf EA, AE, EE}  
      
\end{itemize}    

\caption{Set of rules that treats each fit following conditions from top to bottom and from left to right. The top part of the scheme ($f_{n} \leqslant$ $\varepsilon_f$) only characterises the profiles with stellar component(s), while the second branch ($f_{n} > \varepsilon_f$) also includes the nebular component. In Table \ref{title4}, we merge the categories AN, AAN, and N because the emission is solely due to the nebular component.}
\label{class_scheme}
\end{figure*}

For this ruleset, we define $f_i$ (where $i=1,2,n$) as the integrated flux of Gaussian component $g_i$, which is given by $f_i = A_i \sigma_i \sqrt{2\pi}$. Next we define the following margins:
\begin{align}
\varepsilon_f &= 0.1 F\\
\varepsilon_{\lambda} &= 0.1 \mathrm{ {\AA}}\\
\varepsilon_{\sigma} &= 0.1 \mathrm{ {\AA}}
\end{align}
which, according to our tests, enable the best performance when flagging individual spectra and then assigning distinct morphological categories in the second step of the morphological classification scheme. The flux $f$ is subject to significant variations in absolute value and therefore $\varepsilon_f$ depends on the integrated Gaussian solution $F$ (Equation \eqref{integral}). By doing this, we account for the very strong emissions where Gaussian components, which are relatively weak, but still significant in absolute value, must be weighted appropriately. 

The main purpose of the conditions in Fig. \ref{class_scheme} is to determine whether different fitted Gaussian profiles $g_1, g_2$, and $g_n$ are similar enough in terms of their $\lambda$ and $\sigma$ parameters to represent the same feature (component) of the line profile or if they are so different to indicate two or more distinct components. This can only be evaluated under the limitations of the sampling scheme for the fitted parameters, the resolution of the spectra and their S/N. The defined set of rules and the margins $\varepsilon_f, \varepsilon_{\lambda}$, and $\varepsilon_{\sigma}$ are therefore selected accordingly. The number of spectra that are assigned a certain flag is listed in Table \ref{title4}.

\begin{table}
\captionof{table}{Number of emission type spectra for each of the nine flags.}
\scriptsize
\centering
\begin{tabular}{ c c c c c c c c c }
\bf E & \bf EE & \bf EA & \bf AE & \bf N & \bf EN & \bf EEN & \bf EAN & \bf AEN\\
\hline
348 & 2101 & 1408 & 940 & 1148 & 642 & 938 & 838 & 483\\

\end{tabular}
\label{title4}  
\end{table}

We exclude spectra with flag N from further discussion. Thus we have 990 single intrinsic emission spectra and 6709 spectra with a combination of two intrinsic components, flagged EE(N), EA(N), and AE(N).

\section{SUPPLEMENTARY INFORMATION FROM EXTERNAL SOURCES} \label{sec:addinf}

We extract a collection of data from several sources. As the coordinates of objects are the most reliable search parameters, we use them to retrieve information from SIMBAD, VizieR, and ADS on-line databases. Epoch 2000.0 coordinates of our objects are not identical with those from the catalogues, so we adopt matching limits as described below. Tables \ref{title2} and \ref{title3} provide supplementary information from VizieR, SIMBAD, and ADS databases.

\subsection{Object types from SIMBAD}

We use a search radius of 1 arcsec. Altogether 2546 out of 4459 objects have a SIMBAD match satisfying this criterion. Typical accuracy of the match is $\sim 0.3$~arcsec, implying that the matches we found are secure. We examine the general description of their object type in the SIMBAD database (i.e. the ``otype'' flag). Table \ref{table1} lists the most commonly encountered ``otype'' flags for our targets. While the first two categories are general definitions, those with more than 15 occurrences clearly indicate their nature. These are mostly young, active stars, probably in their early phases of evolution. Some of the objects have been detected as an X-ray or a radio source while, interestingly, there is only a handful of known interacting binary stars. Further examination of objects from our catalogue is thus a promising way to discover several objects belonging to the latter category.

\begin{table}

\small
\captionof{table}{Distribution of objects according to their classification flag ``type'' of the SIMBAD database. N denotes the number of objects from our catalogue that fall into each category.} \label{table1} 
\begin{tabularx}{\linewidth}{c l l l l l l}
\bf N&\bf otype\\
\hline
1025 & star in cluster\\
356 & star\\
244 & young stellar object\\
198 & variable star of Orion type\\
177 & low-mass star (M $<$ M$_{\sun}$)\\
131 & young stellar object candidate\\
85 & pre-main sequence star\\
75 & T Tau-type star\\
73 & pre-main sequence star candidate\\
61 & emission-line star\\
47 & infrared source\\
20 & x-ray source\\
16 & flare star\\
5 & variable star\\
3 & radio source\\
2 & brown dwarf (M $<$ 0.08 M$_{\sun}$)\\
2 & brown dwarf candidate\\
2 & carbon star\\
2 & Cepheid variable star\\
2 & eclipsing binary of W UMa type (contact)\\
2 & Herbig-Haro object\\
2 & open (galactic) cluster\\
2 & Red Giant Branch star\\
2 & variable star of Mira Cet type\\
1 & Be Star\\
1 & dark cloud (nebula)\\
1 & double or multiple star\\
1 & eclipsing binary\\
1 & HII (ionized) region\\
1 & millimetric radio source\\
1 & object of unknown nature\\
1 & rotationally variable star\\
1 & spectroscopic binary\\
1 & sub-millimetric source\\
1 & variable of BY Dra type\\
1 & white dwarf candidate\\
\hline
\end {tabularx}

\end{table}

\subsection{References from the ADS database} \label{adssearch}
 
References from the literature should serve as possible additional information about objects of interest, they are not to be taken as reliable sources of the characteristics of a certain object. 305 out of 4459 objects are matched successfully with references from the ADS database. The search is done based on object coordinates and separately on the main SIMBAD identifier, the latter being justified by the fact that ADS and SIMBAD work together in retrieving the relevant literature. When searching by SIMBAD identifier, the restriction to objects with angular distance $\leqslant 1$ arcsec of matched records in SIMBAD is applied. The results are further filtered by the presence of keywords ``H$\alpha$'', ``emission'', ``objective prism survey'', ``cataclysmic'', ``symbiotic'', ``binary'', ``outburst'', and ``pulsat'' in the title of each paper.

\subsection{VizieR sources}

We perform queries by coordinates to retrieve the properties of our sources contained in VizieR catalogues. Results are presented for each of the following wavelength ranges: Gamma-ray, X-ray, EUV, UV, Optical, IR, and Radio. For 1853 objects in Table \ref{title2}, we list the number of VizieR tables in which a match is found within an angular distance of up to 1 arcsec.

\section{CONTENTS OF THE CATALOGUE} \label{sec:contents}

The catalogue contains a master table with the main results for each analysed spectrum, and its extension for unique objects, which provides supplementary information to the reader. Their content is respectively detailed in Tables \ref{title} and \ref{title2}. Each fitted spectrum is identified through the object name (i.e. the CNAME keyword build from the coordinates of the object) and the date of observation. We give the first five results of the fitting procedure for each spectrum ordered by $\chi^2_r$ in Table \ref{title}, but for all relevant discussion in the main text, only the best result for each spectrum is used.

Some of the objects observed in GES are known to be of peculiar type and are already discussed in the literature or listed in different catalogues. For all the 4459 analysed objects, we present results from the search in ADS, SIMBAD, and VizieR databases in Tables \ref{title2} and \ref{title3}. This search is detailed in Appendix \ref{sec:addinf}. We note that 305 out of 4459 objects are mentioned in the list of ADS references (Table \ref{title3}), and according to SIMBAD, roughly 25\% of 4459 objects already have an indication of H$\alpha$ emission. The electronic version of the catalogue will be made publicly available at the CDS.

\begin{table*}

\small
\captionof{table}{Description of the content for the catalogue of 8846 spectra whose full table will be available at the CDS.} 
\begin{tabularx}{\linewidth}{ l l X }
\bf Label & \bf Unit & \bf Description\\
\hline
DATEOBS & & Date and time of the observation\\
CNAME & & Sexagesimal, equatorial position-based source name in the form: hhmmssss+ddmmsss. Thus the CNAME of an object at 3h 40m 21.s767 and -31$\degr$ 20$\arcmin$ 32.71$\arcsec$ is 03402177-3120327.\\
RA & $\degr$ & RA (J2000)\\
DEC & $\degr$ & DEC (J2000)\\
Halpha\_lambda\_1 & \AA & five comma separated values - first five solutions of the fit ordered by $\chi^2_r$ for the central wavelength of the first Gaussian profile\\
Halpha\_sigma\_1 & \AA & As Halpha\_lambda\_1 but for the $\sigma$ parameter of the first Gaussian profile\\
Halpha\_peak\_1 & & As Halpha\_lambda\_1 but for the height of the peak of the first Gaussian profile in units of normalised flux\\
Halpha\_lambda\_2 & \AA & As Halpha\_lambda\_1 but for the central wavelength of the second Gaussian profile\\
Halpha\_sigma\_2 & \AA & As Halpha\_lambda\_1 but for the $\sigma$ parameter of the second Gaussian profile\\
Halpha\_peak\_2 & & As Halpha\_lambda\_1 but for the height of the peak of the second Gaussian profile in units of normalised flux\\
Halpha\_lambda\_neb & \AA & As Halpha\_lambda\_1 but for the central wavelength of the nebular Gaussian profile\\
Halpha\_sigma\_neb & \AA & As Halpha\_lambda\_1 but for the $\sigma$ parameter of the nebular Gaussian profile\\
Halpha\_peak\_neb & & As Halpha\_lambda\_1 but for the height of the peak of the nebular Gaussian profile in units of normalised flux\\
Halpha\_neb\_indication & & 0 - no indication of nebular emission, 1 - presence of [N~II] emission in object spectrum, 2 - presence of H$\alpha$ and [N~II] emission in sky spectra (see Appendix \ref{gauss_fitting})\\
Halpha\_red\_chi & & five comma separated values - $\chi^2_r$ for first five solutions of the fit\\
Halpha\_red\_chi\_f & & five comma separated values - $\chi^2_{rf}$ for first five solutions of the fit\\
Halpha\_flag & & Flag of spectrum obtained from our ruleset in Fig. \ref{class_scheme} (see text)\\
Halpha\_category & & Category obtained from our classification scheme in Table \ref{phys_cat} (see text)\\
Halpha\_intrinsic & & 1 for intrinsic stellar emission, otherwise NULL (see text)\\

\hline
\end {tabularx}

\label{title} 

\end{table*}

\begin{table*}
\small
\captionof{table}{Description of the content for the supplementary information of 4559 sources whose full table will be available at the CDS.} 
\begin{tabularx}{\linewidth}{ l l X }
\bf Label & \bf Unit & \bf Description\\
\hline
CNAME & & as CNAME in Table \ref{title}\\
RA & $\degr$ & as RA in Table \ref{title}\\
DEC & $\degr$ & as DEC in Table \ref{title})\\
Halpha\_profile\_change & & 0 for no temporal change in classification category of spectra from repeated observations, otherwise NULL (see text)\\
SIMBAD\_main\_id & & Main ID of the source in SIMBAD found for a given CNAME, otherwise NULL\\
SIMBAD\_angular\_distance & arcsec & Angular distance of the object to the source SIMBAD\_main\_id, otherwise NULL\\
SIMBAD\_identifiers & & All SIMBAD identifiers for the source SIMBAD\_main\_id, otherwise NULL\\
SIMBAD\_otype & & SIMBAD object type for the source SIMBAD\_main\_id, otherwise NULL\\
SIMBAD\_stype & & SIMBAD spectral type for the source SIMBAD\_main\_id, otherwise NULL\\
ADS\_literature & & A comma-separated list of indices which correspond to the first column of Table \ref{title3} where we list the ADS references ordered by the number of occurrences for all analysed objects, otherwise NULL\\
VizieR\_n\_Radio & & Number of VizieR tables for Radio wavelength range in which object with CNAME has a match, otherwise NULL\\
VizieR\_n\_IR & & As VizieR\_n\_Radio but for IR wavelength range\\
VizieR\_n\_optical & & As VizieR\_n\_Radio but for optical wavelength range\\
VizieR\_n\_UV & & As VizieR\_n\_Radio but for UV wavelength range\\
VizieR\_n\_EUV & & As VizieR\_n\_Radio but for EUV wavelength range\\
VizieR\_n\_Xray & & As VizieR\_n\_Radio but for X-ray wavelength range\\
VizieR\_n\_Gammaray & & As VizieR\_n\_Radio but for Gamma-ray wavelength range\\
\hline
\end {tabularx}
\label{title2} 
\end{table*}

\begin{table*}
\captionof{table}{Reference list (98 rows, full table will be available at the CDS). We only list the references with N$_{occur} \geqslant$ 50, which denotes the number of objects from the catalogue that have corresponding article in the search results from ADS (see appendix \ref{adssearch}).}
\small
\begin{tabularx}{\linewidth}{c l X l}
\small
\bf id& \bf N$_{occur}$& \bf title& \bf reference\\
\hline
1&157&H$\alpha$ Emission-Line Stars in Molecular Clouds. I. The NGC 2264 Region&\citet{2004AJ....127.1117R}\\
2&98&Catalogue of H-alpha emission stars in the Northern Milky Way&\citet{1999AAS..134..255K}\\
3&98&Catalogue of stars in the Northern Milky Way having H-alpha in emission.&\citet{1997AAHam..11.....K}\\
4&67&Primordial Circumstellar Disks in Binary Systems: Evidence for Reduced Lifetimes&\citet{2009ApJ...696L..84C}\\
5&66&Emission-Line Stars Associated with the Nebulous Cluster NGC 2264.&\citet{1954ApJ...119..483H}\\
6&63&New H-alpha-emission stars in Monoceros OB1 and R1 associations&\citet{1984PASJ...36..139O}\\
\hline
\end{tabularx}
\label{title3} 

\end{table*}

\end{appendices}

\end{document}